\PassOptionsToPackage{table}{xcolor}
\documentclass[acmsmall,authorversion,screen]{acmart}

\usepackage{algorithmic}
\usepackage{graphicx}
\usepackage{subcaption}
\usepackage{textcomp}
\usepackage{pifont}
\usepackage{hyperref}
\usepackage{booktabs}
\usepackage{xspace}
\usepackage{tcolorbox}
\usepackage{tabularx}
\usepackage{threeparttable}
\usepackage{siunitx}
\usepackage{listings}
\usepackage{colortbl}

\usepackage{enumitem}

\definecolor{lightgray}{gray}{0.9}

\newlist{tableitemize}{itemize}{1}
\setlist[tableitemize,1]{label={--},nosep, noitemsep, leftmargin=*, topsep=0pt, parsep=0pt, partopsep=0pt, before=\vspace{-0.6\baselineskip}, after=\vspace{-0.6\baselineskip}}

\newcommand{\yes}{\checkmark}
\newcommand{\no}{\ding{55}} 

\newcommand{\captionheadline}[1]{\textbf{#1}}

\usepackage[english]{babel} 
\usepackage{babel}
\addto\extrasenglish{%

}

\AtBeginDocument{%
  \providecommand\BibTeX{{%
    \normalfont B\kern-0.5em{\scshape i\kern-0.25em b}\kern-0.8em\TeX}}}

\microtypecontext{spacing=nonfrench}

\usepackage[acronym,nohypertypes={acronym},shortcuts,nonumberlist,nolist]{glossaries} 
\newacronym{AAS}{AAS}{Account Adoption Scale}
\newacronym{ATI}{ATI}{Affinity for Technology Interaction}
\newacronym{BLE}{BLE}{Bluetooth Low Energy}
\newacronym{COTS}{COTS}{commercial off-the-shelf}
\newacronym{E2EE}{E2EE}{end-to-end encrypted}
\newacronym{CTAP}{CTAP}{Client to Authenticator Protocol}
\newacronym[description={Deutsches Forschungsnetz ("German research network")}]{DFN}{DFN}{Deutsches Forschungsnetz}
\newacronym{DNS}{DNS}{Domain Name System}
\newacronym{FIDO}{FIDO}{Fast IDentity Online Alliance}
\newacronym{ASK}{ASK}{amplitude-shift keying}
\newacronym{FSK}{FSK}{frequency-shift keying}
\newacronym{MFSK}{MFSK}{multiple frequency-shift keying}
\newacronym{PSK}{PSK}{phase-shift keying}
\newacronym{DSSS}{DSSS}{direct-sequence spread spectrum}
\newacronym{OFDM}{OFDM}{orthogonal frequency-division multiplexing}
\newacronym{SSB}{SSB}{single-sideband modulation}
\newacronym{AWGN}{AWGN}{additive white gaussian noise}
\newacronym{GDPR}{GDPR}{General Data Protection Regulation}
\newacronym{HTTPS}{HTTPS}{Hypertext Transfer Protocol Secure}
\newacronym{IP}{IP}{Internet Protocol}
\newacronym{IoT}{IoT}{Internet of Things}
\newacronym{ISO}{ISO}{International Organization for Standardization}
\newacronym{NIST}{NIST}{National Institute of Standards and Technology}
\newacronym{OTSP}{OTSP}{Online Token Status Protocol}
\newacronym{PAS}{PAS}{Participant Adoption Scale}
\newacronym{SMS}{SMS}{Short Message Service}
\newacronym{SUS}{SUS}{System Usability Scale}
\newacronym{TLS}{TLS}{Transport Layer Security}
\newacronym{TUDA}{TUDA}{Technische Universität Darmstadt}
\newacronym{U2F}{U2F}{Universal 2nd Factor}
\newacronym{UAF}{UAF}{Universal Authentication Framework}
\newacronym{USB}{USB}{Universal Serial Bus}
\newacronym{W3C}{W3C}{World Wide Web Consortium}
\newacronym{WebAuthn}{WebAuthn}{Web Authentication}

\newacronym[description={One-Factor Authentication}]{1FA}{1FA}{one-factor authentication}
\newacronym[description={Two-Factor Authentication}]{2FA}{2FA}{two-factor authentication}
\newacronym[description={Application Programming Interface}]{API}{API}{application programming interface}
\newacronym[description={Application}]{app}{app}{application}
\newacronym[description={Certification Authority}]{CA}{CA}{certification authority}
\newacronym[description={Effect Size}]{ES}{ES}{effect size}
\newacronym[description={False Discovery Rate}]{FDR}{FDR}{false discovery rate}
\newacronym[description={False Rejection Rate}]{FRR}{FRR}{false rejection rate}
\newacronym[description={Graphical User Interface}]{GUI}{GUI}{graphical user interface}
\newacronym[description={radio frequency}]{RF}{RF}{radio frequency}
\newacronym[description={Institutional Review Board}]{IRB}{IRB}{institutional review board}
\newacronym[description={Internet Service Provider}]{ISP}{ISP}{internet service provider}
\newacronym[description={Multi-Factor Authentication}]{MFA}{MFA}{multi-factor authentication}
\newacronym[description={Machine-in-the-Middle}]{MitM}{MitM}{machine-in-the-middle}
\newacronym[description={Near-Field Communication}]{NFC}{NFC}{near-field communication}
\newacronym[description={Operation System}]{OS}{OS}{operating system}
\newacronym[description={One-Time Password}]{OTP}{OTP}{one-time password}
\newacronym[description={Out-Of-Band}]{OOB}{OOB}{out-of-band}
\newacronym[description={Personal Computer}]{PC}{PC}{personal computer}
\newacronym[description={Personal Identification Number}]{PIN}{PIN}{personal identification number}
\newacronym[description={Public-Key Infrastructure}]{PKI}{PKI}{public-key infrastructure}
\newacronym[description={Quick-Response}]{QR}{QR}{quick-response}
\newacronym[description={Secure Device Pairing}]{SDP}{SDP}{secure device pairing}
\newacronym[description={Service Set Identifier}]{SSID}{SSID}{service set identifier}
\newacronym[description={Single Sign-On}]{SSO}{SSO}{single sign-on}
\newacronym[description={Transaction Authentication Number}]{TAN}{TAN}{transaction authentication number}
\newacronym[description={Trust on First Use}]{TOFU}{TOFU}{trust on first use}
\newacronym[description={Trusted Platform Module}]{TPM}{TPM}{trusted platform module}
\newacronym[description={Uniform Resource Locator}]{URL}{URL}{uniform resource locator}
\newacronym[description={Wireless Local Area Network}]{WLAN}{WLAN}{wireless local area network}
\newacronym{WoS}{WoS}{Web of Science}
\newacronym{MIMO}{MIMO}{multiple-input and multiple-output}
\newacronym{TER}{TER}{total error rate}
\newacronym{TX}{TX}{transmitter}
\newacronym{RX}{RX}{receiver}

\AtBeginDocument{%
  \providecommand\BibTeX{{%
    Bib\TeX}}}

\setcopyright{acmlicensed}

\acmJournal{TIOT}
\acmYear{2026} \acmVolume{7} \acmNumber{1} \acmArticle{8} \acmMonth{2}\acmDOI{10.1145/3779439}

\acmSubmissionID{1874}

\begin{document}

\title[Evaluating Acoustic Data Transmission Schemes for Ad-Hoc Communication Between Nearby Smart Devices]{Evaluating Acoustic Data Transmission Schemes for Ad-Hoc Communication Between Nearby Smart Devices}

\author{Florentin Putz}
\orcid{0000-0003-3122-7315}
\email{fputz@seemoo.de}
\author{Philipp Fortmann}
\orcid{0009-0005-6544-6876}
\email{pfortmann@seemoo.de}
\author{Jan Frank}
\orcid{0009-0003-7459-1624}
\email{jfrank@seemoo.de}
\author{Christoph Haugwitz}
\orcid{0000-0001-6756-9988}
\email{christoph.haugwitz@tu-darmstadt.de}
\author{Mario Kupnik}
\orcid{0000-0003-2287-4481}
\email{mario.kupnik@tu-darmstadt.de}
\author{Matthias Hollick}
\orcid{0000-0002-9163-5989}
\email{mhollick@seemoo.de}
\affiliation{%
  \institution{Technical University of Darmstadt}
  \city{Darmstadt}
  \country{Germany}
  }

\renewcommand{\shortauthors}{Putz et al.}

\begin{abstract}
Acoustic data transmission offers a compelling alternative to Bluetooth and NFC by leveraging the ubiquitous speakers and microphones in smartphones and IoT devices.
However, most research in this field relies on simulations or limited on-device testing, which makes the real-world reliability of proposed schemes difficult to assess.
We systematically reviewed 31 acoustic communication studies for commodity devices and found that none provided accessible source code.
After contacting authors and re-implementing three promising schemes, we assembled a testbed of eight representative acoustic communication systems.
Using over \num{11000} smartphone transmissions in both realistic indoor environments and an anechoic chamber, we provide a systematic and repeatable methodology for evaluating the reliability and generalizability of these schemes under real-world conditions.
Our results show that many existing schemes face challenges in practical usage, largely due to severe multipath propagation indoors and varying audio characteristics across device models.
To support future research and foster more robust evaluations, we release our re-implementations alongside the first comprehensive dataset of real-world acoustic transmissions.
Overall, our findings highlight the importance of rigorous on-device testing and underscore the need for robust design strategies to bridge the gap between simulation results and reliable IoT deployments.
\end{abstract}

\ccsdesc[500]{Networks~Mobile ad hoc networks}
\ccsdesc[100]{Networks~Short-range networks}
\ccsdesc[100]{Networks~Network reliability}
\ccsdesc[500]{Networks~Network performance evaluation}
\ccsdesc[500]{Human-centered computing~Ubiquitous and mobile devices}
\ccsdesc[500]{Human-centered computing~Smartphones}
\ccsdesc[100]{Hardware~Signal processing systems}

\keywords{acoustic communication; ultrasonic communication; near-ultrasonic; audio-based communication; data-over-sound; reproducibility; conceptual replication; generalizability; dataset; open data; open source; audio recordings; iot; ubiquitous computing; experimental testbed; systematic literature study; re-implementation; performance evaluation; device-to-device communication; smartphone-to-smartphone communication; short-range; ad-hoc; interoperability; spontaneous; nearby; proximity}


\maketitle

\section{Introduction}
Smartphones and \gls{IoT} devices can communicate using sound waves via their built-in speakers and microphones~\cite{lopesAcousticModemsUbiquitous2003}.
Acoustic data transmission promises to be a practical alternative to Bluetooth and \gls{NFC}, given the universal presence of audio I/O in devices like smartphones, smartwatches, smart speakers, and home automation systems. 
Over the past decade, there has been increasing interest in using acoustic communication as a wireless, short-range ad-hoc channel to facilitate ubiquitous and mobile computing use cases, such as mobile payments \cite{wang2021chirpmu}, context-aware computing~\cite{madhavapeddy2003ContextAware,madhavapeddyAudioNetworkingForgotten2005}, or
location-based services \cite{zhang2019priwhisper,nandakumar2013dhwani}.
Its location-limited nature also fits well with ubiquitous computing scenarios, where users intuitively bring devices into proximity so they can 
``talk''~\cite{gerasimovThingsThatTalk2000},  aligning with intuitive user behaviour~\cite{chong2013How,jokela2014FlexiGroups,jokela2015Connecting}.
The widespread availability of low-cost \gls{IoT} hardware equipped with speakers and microphones has even led to commercial uptake, 
such as the smart-speaker manufacturer Sonos employing acoustic communication for spatial awareness and device pairing \cite{sonos2022near}.

Despite the growing body of research proposing acoustic data transmission systems \cite{lopes2001aerial,lee2015chirp,getreuer2018ultrasonic,cai2022boosting,zhang2014priwhisper,goncalves2017acoustic}, many have been tested only through simulations or under very limited real-world conditions.
Real-device testing, however, is crucial to account for factors like hardware imperfections, user handling, realistic signal propagation, and background noise---factors that are notoriously difficult to model precisely.
Moreover, acoustic channels differ substantially from radio channels: 
sound waves travel more than \num{87000} times slower than electromagnetic waves, leading to delay spreads on the order of tens of milliseconds \cite{shi2023longrange,lee2015chirp}.
Acoustic signals also suffer from highly variable ambient noise and, due to low sample rates on most smartphones and \gls{IoT} devices, bandwidth is limited to at most \SI{22}{kHz}, making high-throughput transmissions challenging.
Existing evaluations typically focus on individual devices or scenarios, providing little insight into how well a scheme might generalize across diverse real-world settings.

At the same time, there is a distinct gap between academic research and commercial practice.
Commercial products for acoustic data transmission on smart devices typically achieve only \SIrange{10}{100}{bps} \cite{cueaudioinc.2024cue,getreuer2018ultrasonic,sonos2022near} and occasionally up to \SI{200}{bps} \cite{stimshop2024wius}, prioritizing robustness and reliability in noisy environments. In contrast, researchers often report much higher throughput up to \SIrange{500}{10000}{bps} \cite{zhang2014priwhisper,cai2022boosting,goncalves2017acoustic} or beyond \cite{yamamoto202232kbps}. This discrepancy raises fundamental questions about whether these systems can maintain reliability in real-world scenarios, or whether additional constraints---such as hardware limitations, user behavior, or environmental interference---ultimately force practical systems to operate at lower data rates.

A major barrier to resolving these questions is that publications in this field often omit source code, limiting independent validation and hindering follow-up work.
In this paper, we take a first step toward consolidating existing findings by investigating how well representative acoustic data transmission schemes generalize to various practical scenarios. Specifically, we address the following research questions:\\

\noindent\textbf{RQ1:} ``How challenging is it to obtain or re-implement systems proposed in this field?''\\[0.4em]
\noindent\textbf{RQ2:} ``How well do these schemes generalize to ad-hoc smartphone communication use cases?''\\[0.4em]
\noindent\textbf{RQ3:} ``Which practical challenges affect nearby acoustic data transmission between smart devices?''

\subsection{Contributions}
This paper presents the first \textbf{independent evaluation of the generalizability of acoustic data communication schemes} proposed by other researchers, comparing them in realistic settings. Our contributions are as follows:

\begin{itemize}
    \item \textbf{Systematic literature study.} We conducted a systematic review and identified 31 publications proposing acoustic data transmission systems for \gls{COTS} devices (\autoref{sec:literature-study}). Regrettably, none of these studies provide implementation source code.
    \item \textbf{Sourcing and re-implementation of representative schemes.} We reached out to the authors for software implementations. Only three authors supplied functioning code, one provided a non-functioning implementation, and the others either did not respond or had lost access. To better understand the generalizability of acoustic communication research, we additionally re-implemented three promising schemes
    (\autoref{sec:obtaining-implementations}).
    \item \textbf{Evaluation of generalizability.} In total, we evaluated the practical reliability of eight acoustic communication schemes \cite{lee2015chirp,lopes2001aerial,getreuer2018ultrasonic,cai2022boosting,zhang2014priwhisper,goncalves2017acoustic}, including two variants of the popular open-source project ggwave \cite{gerganov2024ggwave} (\autoref{sec:evaluation}).
    Our testbed focuses on diverse smartphone models as a practical and accessible subset of IoT devices, although the underlying insights hold for any device with comparable audio input and output capabilities.
    We expand prior work by evaluating these schemes across various IoT use cases and analyzing how factors such as device distance, model, user handling, and background noise affect communication reliability (measured by the rate of errors in transmissions).
    \item \textbf{Measurement dataset.} We publish our re-implementations and analysis code along with the first dataset in this field---\num{11900} real recordings of acoustic data transmissions---to support transparent, reproducible research \cite{replicationpackage}.
\end{itemize}

\subsection{Scope}
This work supports the IoT research community by analyzing acoustic data transmission schemes from the perspective of researchers or developers who want to integrate this technology into smart devices for nearby data exchange. From this perspective, acoustic communication should be as seamless to adopt as Bluetooth or \gls{NFC}, requiring no specialized knowledge of the underlying wireless standards.
Accordingly, we evaluate representative schemes on multiple smartphone models, which serve as the most common and versatile IoT form factor.
Rather than focusing on physical-layer performance or simulations, we treat the complete communication system as a black box, including all components---such as synchronization and error correction---as originally described by the authors, ensuring we analyze its practical suitability on real devices.

\section{System model}\label{sec:system-model}
\begin{figure}[h]
\centering
\includegraphics[page=1,width=\columnwidth]{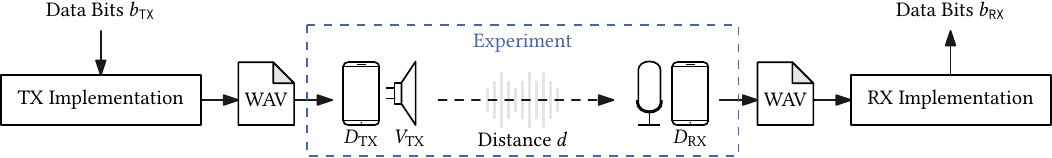}
\caption{\captionheadline{Acoustic data transmission model.} In our experiments, we play the WAV files from the TX implementation on device $D_{\text{TX}}$ (at volume $V_{\text{TX}}$) and record this on device $D_{\text{RX}}$. The resulting WAV file gets decoded using the RX implementation.}
\Description{Block diagram of data flow and experimental setup. Rectangular blocks depict a complete transmitter–channel–receiver chain. On the far left, an arrow labeled ``Data Bits $b_\mathtt{TX}$'' enters a block labeled “TX Implementation.” From this block, an arrow labeled ``WAV'' points to a central block labeled ``Experiment''. Inside this middle block is the transmitting smartphone $D_{\text{TX}}$ with speaker $V_{\text{TX}}$, at a distance $d$ from the receiving smartphone $D_{\text{RX}}$ and it's microphone.
This middle block indicates the physical configuration and playback volume used during the experiment. A second arrow labeled ``WAV'' leaves the experiment block and enters a block on the right labeled ``RX Implementation''. From this final block, an arrow labeled ``Data Bits $b_\mathtt{RX}$'' exits, representing the recovered data bits at the receiver.}
\label{fig:model}
\end{figure}

This study evaluates \textit{acoustic data transmission schemes}: software implementations that broadcast data from one device to nearby devices using sound waves.
From a system developer's perspective, each scheme acts as a black box that transmits data on one device and receives it on another.
For our evaluation, we model this capability as an interface with two functions, depicted in \autoref{fig:model}:

\begin{enumerate}
    \item A \textbf{transmitter}, $\mathtt{wav} = \mathtt{TX}(b_\mathtt{TX})$, which converts data bits into a WAV file for playback on a loudspeaker.
    \item A \textbf{receiver}, $b_\mathtt{RX} = \mathtt{RX}(\mathtt{wav})$, which extracts data from a WAV file containing the microphone recording of the transmission.
\end{enumerate}

This interface separates the digital signal processing of the modulator and demodulator from audio playback and recording. Although some schemes handle playback and recording internally, we treat them independently for clarity.
In practice, a developer might not store intermediate WAV files but instead stream samples directly from a buffer.
Nonetheless, our interface fundamentally reflects real-world usage and facilitates a fair comparison of each scheme under uniform playback and recording conditions.

\subsection{Scenarios}\label{sec:use-cases}

We focus on the use case of data exchange between nearby smart devices, where software modifications are possible (e.g., by installing apps on smartphones) but changes to hardware or firmware are not.
Our emphasis is on broadcast schemes that enable communication without requiring special arrangements with the communication partner.
We divide common use cases for acoustic communication (\autoref{sec:bg-rw}) into three categories based on the communication distance---a key factor affecting usability.
The practical difference in usability is significant, ranging from the need for devices to touch (as in \gls{NFC}) to the flexibility of holding your device several meters away from your communication partner.
Throughout this paper, we focus on these concrete communication scenarios, which represent most current use cases:

\begin{enumerate}
    \item \textbf{Near distance} ($d \leq \SI{10}{\centi\meter}$): Transmitting a public key or a small file ($\approx$\SI{4096}{\bit})
    over a short distance, similar to \gls{NFC}. Examples are mobile payments \cite{copsonictechnologies2024copsonic,theamericanbusinessawards2014verifone}, secure device pairing \cite{zhang2014priwhisper,kumparak2013slicklogin,trillbitinc.2024trillbit}, and access control \cite{nandakumar2013dhwani,copsonictechnologies2024copsonic}.

    \item \textbf{Medium distance} ($\SI{10}{\centi\meter} < d \leq \SI{1}{\meter}$): Transmitting a public key fingerprint or a hash value ($\approx$\SI{128}{\bit})
    over a medium distance. Examples are \gls{IoT} device provisioning \cite{sonos2022near,scheck_2023_speaker_pairing_demo}, side channels for air-gapped devices \cite{sadeq2022privacy}, and control channels for personal electronic devices and hearing aids \cite{miegel2018wireless}.

    \item \textbf{Far distance} ($d > \SI{1}{\meter}$): Transmitting a short ID ($\approx$\SI{16}{\bit})
    over larger distances. Examples are providing location-specific information near public signs or in shops \cite{dey2001understanding}, indoor \gls{IoT} communication \cite{haus2019enhancing,ji2020authenticating}, and enhancing media experiences, such as social television \cite{lin2017tagscreen} and second screen services via hidden audio data in TV and radio \cite{ka2016nearultrasound}.
\end{enumerate}

\section{Related work}\label{sec:bg-rw}

In this section, we briefly review prior research on acoustic data transmission before presenting our systematic literature study in \autoref{sec:literature-study}.

Although smartphone and \gls{IoT} audio hardware is primarily intended for media playback and voice communication, it can also be repurposed for sensing~\cite{wang2018cfmcw,chen2022swipepass,zhao2023ultrasnoop} or even for aerial acoustic data transmission. In the latter case, speakers encode information in sound waves using frequency or amplitude variations, similar to electromagnetic communication~\cite{lopesAcousticModemsUbiquitous2003}, thereby enabling data exchange across a diverse group of devices, including smartphones and voice-controlled appliances, without special \gls{RF} hardware~\cite{gerasimovThingsThatTalk2000}.
This communication channel is limited by the typical \SIrange{0}{22}{kHz} audio range (due to common sampling rates of \SI{44.1}{kHz}), pronounced frequency selectivity \cite{hromadova2022frequency}, and non-linearities at higher volumes (see \autoref{sec:practical-challenges}).
Data can be transmitted audibly or using near-ultrasonic frequencies, and---unlike Wi-Fi or Bluetooth---no network setup  or pairing is required, allowing ad-hoc broadcasts over several meters.
Because it is entirely software-defined, acoustic communication can be easily deployed on existing devices and tailored to support a variety of communication needs~\cite{lopes2001aerial}.

Research in ubiquitous and mobile computing has explored acoustic data transmission for context-aware computing~\cite{madhavapeddy2003ContextAware,madhavapeddyAudioNetworkingForgotten2005} and location-limited communication between nearby devices~\cite{arentz2011ultrasonic,wang2022motorbeat}. The limited range has also attracted interest for secure device interactions~\cite{ji2020authenticating,putz2024soundsgood}, including device pairing~\cite{putz2020acoustic,goodrich2006Loud,goodrich2009Using,zhang2014priwhisper,putz2024pairsonicdemo} and protecting implantable medical devices~\cite{halperinPacemakersImplantableCardiac2008}. User experience aspects of acoustic transmissions have been evaluated~\cite{mehrabi2019Evaluating,lopes2001aerial,lopesAcousticModemsUbiquitous2003,putz2024soundsgood}, as the audibility of the data exchange is a new aspect distinguishing it from previous wireless technologies. Various commercial applications leverage acoustic data transmission for mobile payments \cite{theamericanbusinessawards2014verifone,copsonictechnologies2024copsonic}, context-aware applications~\cite{lisnrinc.2024radius}, location-based services~\cite{cueaudioinc.2024cue,shopkickinc.2024shopkick}, data transfer~\cite{stimshop2024wius}, and device pairing~\cite{kumparak2013slicklogin,trillbitinc.2024trillbit}.
Despite its application potential, no previous work has attempted to systematically evaluate these systems on real devices. The absence of published source code hinders replication of prior results.

\section{Literature study}\label{sec:literature-study}
We conducted a systematic literature survey to gather all papers that propose new aerial acoustic software-defined communication systems, according to the following scope.

\subsection{Scope}
To ensure our research aligns with ubiquitous computing and everyday consumer interactions, we focused our literature survey on papers that describe practical implementations of aerial acoustic device-to-device communication specifically for \gls{COTS} devices, like smartphones.
We excluded studies related to other types of acoustic communication, such as underwater or industrial uses, and those that are not applicable to general-purpose data transmission systems.

\subsection{Method}
To identify relevant literature on acoustic data transmission schemes, we began by collecting a wide array of papers related the subject.
We then refined this collection to include only those with research implementations for our target use case. Our method was guided by PRISMA \cite{page2021prisma}, which is a framework for assessing large bodies of medical research in a structured manner. PRISMA provides a flowchart with three steps:

\begin{enumerate}
    \item \textbf{Identification:} Define the data sources and search strategy.
    \item \textbf{Screening:} Filter the research according to the desired scope.
    \item \textbf{Inclusion:} Select which papers to include.
\end{enumerate}

\subsubsection{Identification}
As the information source for this literature study we used the \gls{WoS},\footnote{\url{https://clarivate.com/products/web-of-science/}, last access: 2025-03-03} which is a comprehensive database indexing journals and conference proceedings.
We selected \gls{WoS} due to its extensive, regularly updated catalog and its advanced filtering capabilities with nested logical operators, which are crucial for narrowing the results to our specific scope.
Our search strategy balanced breadth and manageability. We designed and executed two mutually exclusive search queries\footnote{Search query 1: \url{https://www.webofscience.com/wos/woscc/summary/4c89e96d-5de5-4543-9167-23d73086d8e6-8a0aeac3/relevance/1}; search query 2: \url{https://www.webofscience.com/wos/woscc/summary/392d145e-a4ac-4aff-8995-ad2103dd116b-898dca59/relevance/1}.} on May 18, 2023, yielding \SI{631} papers. These queries targeted papers published between 2000 and 2022 in the fields of acoustics, computer science, and engineering, focusing on relevant keywords in the titles. We excluded papers related to nautical or medical fields by blocking specific keywords associated with underwater communication and ultrasonic imaging. The second query broadened the search to include a wider range of keyword combinations. We refer to the query URLs\footnotemark[\value{footnote}] for details and keywords.

\subsubsection{Screening}
We processed all identified papers to determine their relevance to our scope.
In the first iteration, we reviewed the titles and abstracts of all \num{631} papers and eliminated \num{422} papers that were clearly out of scope---such as those focusing on speech-based communication, tracking with audio, or communication through metal barriers.
In a second iteration, we examined the full text of the remaining \num{209} papers, classifying them based on their applicability to ubiquitous computing use cases and whether they describe a testable implementation for \gls{COTS} devices.
We excluded papers dependent on specialized hardware, like microphone or speaker arrays for \gls{MIMO} communication, and those embedding data in music, unless the scheme could be utilized as a general-purpose communication system.

\subsubsection{Inclusion}
In total, our literature study identified \num{31} papers that directly align with our study's scope and describe a testable communications scheme (as shown in \autoref{tab:survey}).

\begin{table}[bt]
\caption{\captionheadline{Surveyed acoustic communication schemes,} sorted by throughput. The right table continues from the left. Data not available in the publications is marked with ``?''.}
\label{tab:survey}
\begin{threeparttable}
\begin{minipage}[t]{.48\textwidth}
\footnotesize
\begin{tabular}{@{}l@{\hspace{0.5em}}l@{\hspace{0.6em}}r@{\hspace{1em}}c@{\hspace{0.3em}}r@{}}
\toprule
\textbf{Name} & \textbf{Modulation} & \rotatebox{90}{\textbf{Frequencies} [kHz]} & \rotatebox{90}{\textbf{Inaudible}} & \rotatebox{90}{\textbf{Throughput} [bps]} \\ 
\midrule
Moriyama et al.~\cite{moriyama2017creation} & ASK & \textasciitilde 19 & \yes & 5  \\
Kim et al.~\cite{kim2019dataoversound}& 16-FSK & 18\,--\,20 & \yes & 5 \\
Ka et al.~\cite{ka2016nearultrasound}& Chirp & 18.5\,--\,19.5 & \yes & 15  \\
Lee2015 \cite{lee2015chirp}& Chirp & 19.5\,--\,22 & \yes & 16  \\
Lee2020 \cite{lee2020reliable} & Chirp & 18.5\,--\,19.5 & \yes & 16  \\
Hornych et al.~\cite{hornych2020nearultrasonic}& ASK, FSK & 18\,--\,20 & \yes & 33  \\
TagScreen \cite{lin2017tagscreen}& Chirp & 18\,--\,20 & \yes & 36  \\
Haus et al.~\cite{haus2019enhancing}& OFDM + ASK & 22\,--\,24 & \yes & 64  \\
Miegel et al.~\cite{miegel2018wireless}& OOK, FSK & 16\,--\,20 & \yes & 84  \\
Rustle \cite{jin2022furtively}& \textit{custom}\tnote{1} & 0\,--\,15 & \no & 90  \\
Nearby \cite{getreuer2018ultrasonic} & DSSS + MFSK & 18.5\,--\,20 & \yes & 95  \\
Bang et al.~\cite{bang2016data}& FSK & 18\,--\,22 & \yes & 170  \\
Sadeq et al.~\cite{sadeq2022privacy}& FSK & 18\,--\,19 & \yes & 440  \\
HRCSS \cite{cai2022boosting}& Chirp & 18\,--\,22 & \yes & 500  \\
Dolphin \cite{wang2016messages,zhou2019dolphin} & OFDM + ASK & 8\,--\,20 & \no & 500  \\
\bottomrule
\end{tabular}
\end{minipage}\hspace{1.5em}
\begin{minipage}[t]{.48\textwidth}
\footnotesize
\begin{tabular}{@{}l@{\hspace{0.5em}}l@{\hspace{0.6em}}r@{\hspace{1em}}c@{\hspace{0.5em}}r@{}}
\toprule
\textbf{Name} & \textbf{Modulation} & \rotatebox{90}{\textbf{Frequencies} [kHz]} & \rotatebox{90}{\textbf{Inaudible}} & \rotatebox{90}{\textbf{Throughput} [bps]} \\ 
\midrule
Gon\c{c}alves et al.~\cite{goncalves2017acoustic} & OFDM + PSK & 0\,--\,4 & \no & 600 \\
PriWhisper \cite{zhang2014priwhisper,zhang2019priwhisper} & MFSK & 9\,--\,17 & \no & 1000 \\
DigitalVoices \cite{lopes2001aerial}& ASK, (M)FSK, SS & 1\,--\,10 & \no & 1280 \\
A-NFC \cite{lin2015anfc} & QPSK & 16--21 & \yes & 2200 \\
Dhwani \cite{nandakumar2013dhwani}& OFDM + PSK & 6\,--\,7 & \no & 2400 \\
Hush \cite{novak2019ultrasound}& OFDM + QAM & 17.5\,--\,21  & \yes & 4900 \\
He et al.~\cite{he2018novel}& AM, FSK & 0\,--\,22  & \no & 6000 \\
Yamamoto et al.~\cite{yamamoto202232kbps}& 16-QAM & ? & ? & 32000 \\
Arentz et al.~\cite{arentz2011ultrasonic}& ASK & 20--23  & \yes & ? \\
Bits-over-Sound \cite{kanade2018audiointernet} & DBPSK & ? & \no & ? \\
Jeon et al.~\cite{jeon2016noncoherent}& PSK & 20  & \yes & ? \\
No Entry \cite{kwak2019no}& Chirp & ? & ? & ? \\
Sasano et al.~\cite{sasano2014performance}& OFDM + PSK & 6.4\,--\,8  & \no & ? \\
SilentInformer \cite{satapathy2020silentinformer}& FSK & 17\,--\,19  & \yes & ? \\
&&&&\\
\bottomrule
\end{tabular}
\end{minipage}
  \scriptsize
  \begin{tablenotes}
    \item[1] Rustle's custom approach is called ``furtive modulation''.
  \end{tablenotes}
\end{threeparttable}
\end{table}

\section{Obtaining acoustic data transmission implementations}\label{sec:obtaining-implementations}

To independently evaluate existing acoustic communication systems and embed them in IoT testbeds, we first needed to obtain functional implementations.  
We applied three different methods to acquire the implementations (see \autoref{sec:discussion-obtaining} for a discussion of the associated challenges).

\subsection{Publicly available implementations}
None of the 31 papers surveyed referenced a publicly available implementation.

\subsection{Author requests}\label{sec:contacting-authors}
We contacted every author via email as shown in \autoref{fig:contacting_authors}. Only six replied and just three of those responses provided a functional implementation:

\begin{figure}[tbp]
\centerline{\includegraphics[width=1\columnwidth]{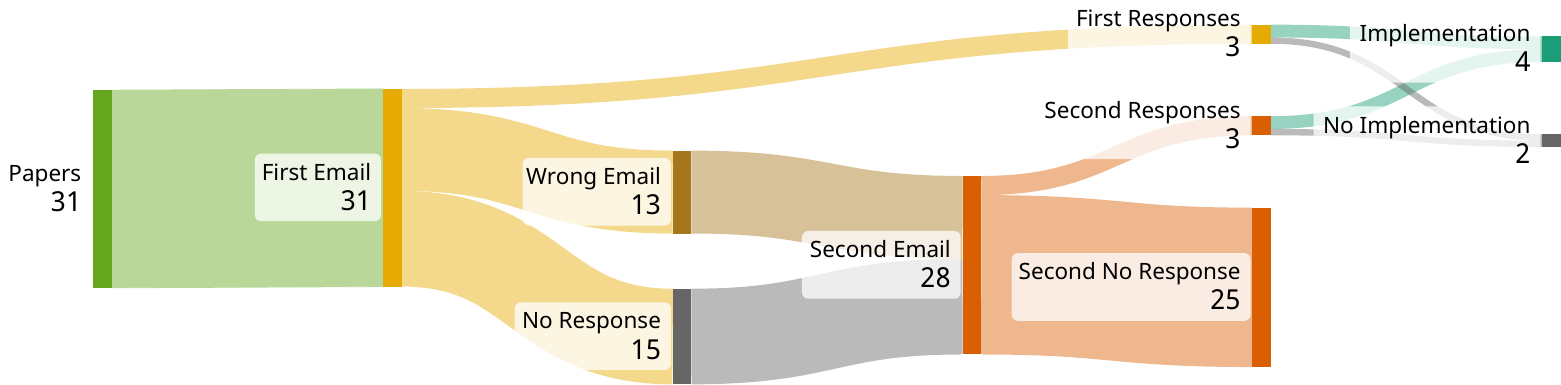}}
\caption{\captionheadline{Responses from authors} when we requested implementations for 31 publications on acoustic data transmission. We contacted the first authors and all co-authors. For 13 papers, the provided email addresses were outdated, prompting us to search for current contact details manually.
We followed up if there was no reply after three weeks.
Of the four implementations received, we evaluated three due to runtime errors in the fourth. The other responding authors reported no longer having access to the implementations, citing reasons such as the lead author departing the university or hardware failures.}
\label{fig:contacting_authors}
\Description{The figure is a Sankey diagram summarizing how many author teams responded when we requested implementations for 31 papers on acoustic data transmission.

On the left, a node labeled ``Papers: 31'' flows entirely into ``First Email: 31'', indicating that an initial email was sent for all 31 papers. This node then splits into three branches: ``First Responses: 3'', ``Wrong Email: 13'', and ``No Response: 15”, showing that only 3 papers yielded an initial reply, 13 had outdated or incorrect email addresses, and 15 produced no reply at all. From there, the diagram shows a follow-up stage labeled ``Second Email: 28'', representing additional contact attempts. This second round of contact leads to ``Second Responses: 3'' and ``Second No Response: 25'', meaning only 3 more responses were obtained while 25 remained unanswered. Finally, among all responses, a smaller flow leads to an ``Implementation: 4'' node and another to ``No Implementation: 2'', indicating that four author teams provided runnable implementations, whereas two responded but reported that their code was no longer available.}
\end{figure}

\begin{itemize}
    \item \textbf{Digital Voices}, proposed by Lopes et al.~\cite{lopes2001aerial} in 2001, is among the earliest documented aerial acoustic communication schemes.
    It uses binary \gls{ASK} across eight carrier frequencies. The authors provided their Java source code.
    \item \textbf{HRCSS}, introduced by Cai et al.~\cite{cai2022boosting}, is a chirp-based communication scheme that uses multiple chirp carriers simultaneously.
    The MATLAB source code provided to us lacks their bidirectional probing feature, so we only evaluate the data transmission.
    \item \textbf{Gon\c{c}alves et al.}~\cite{goncalves2017acoustic} 
    proposed a scheme using \gls{OFDM} and \gls{PSK} on each subcarrier, incorporating a short chirp sequence for synchronization. The authors provided us with the MATLAB source code.
\end{itemize}

\noindent
A fourth response (Hush~\cite{novak2019ultrasound}) was unusable on current Android versions.\footnote{The authors of Hush~\cite{novak2019ultrasound} shared their Android implementation, but we encountered fatal runtime errors that hindered signal transmission and recording. These issues were likely due to changes in the Android ecosystem and incomplete code.
The project has not been updated since 2017 and the authors acknowledged in their correspondence that they are aware of the lack of maintenance. Despite this, we still commend Novak et al.~\cite{novak2019ultrasound} for making their source code available.}

\subsection{Re-implementing acoustic data transmission schemes}\label{sec:reimplementation}

We re-implemented the schemes from three highly cited papers, chosen to represent our three target use cases, each with distinct throughput and maximum transmission distances:

\begin{itemize}
    \item \textbf{Lee et al.} \cite{lee2015chirp} claim to be the first to create a prototype that uses chirps in practical acoustic communication, targeting the far-distance use case. In their system, each communication frame begins with a long chirp preamble, followed by 16 individual chirps.
    \item \textbf{Nearby} was proposed by Getreuer et al.~\cite{getreuer2018ultrasonic} and has been integrated into several Google products, including Google Nearby. It is tailored for the medium-distance use case and employs a combination of \gls{MFSK}, \gls{DSSS}, and \gls{SSB}.
    \item \textbf{PriWhisper}, introduced by Zhang et al.~\cite{zhang2014priwhisper}, is an acoustic communication system for the nearby use case, akin to \gls{NFC}. It utilizes \gls{MFSK} and includes friendly jamming for physical-layer security. However, in our study, we focus solely on the communication capabilities of PriWhisper and have not implemented its security features.
\end{itemize}

\noindent
We validated our re-implementations using simulations and practical experiments, ensuring that they perform similar to their original evaluation results.
The system proposed by Lee et al.~\cite{lee2015chirp} is thoroughly detailed, enabling us to implement all aspects as described in their paper.
The other two required filling in undocumented details (see \autoref{sec:reimplementation-challenges}).
We publish our MATLAB re-implementations as part of our replication package \cite{replicationpackage}.

\begin{table}[t]
\centering
\caption{\captionheadline{Evaluated acoustic communication schemes}, sorted by throughput. The first column specifies the source of the implementation: P -- referenced in the paper; A -- by contacting the authors; R -- by re-implementing them ourselves; N -- non-academic solution for comparison. In the last three columns, we classify each scheme's a priori suitability for the three use case categories defined in \autoref{sec:use-cases}, based on their specified throughput and maximum distance.}
\label{tab:schemes}
\small
\begin{threeparttable}
\begin{tabular}{@{}c ll@{\hspace{0.4em}}r@{\hspace{0.5em}}c@{\hspace{0.5em}}r|c@{\hspace{0.5em}}c@{\hspace{0.5em}}c@{}}
\toprule
\textbf{Via} & \textbf{Name} & \textbf{Modulation} & \textbf{Frequencies} & \textbf{Inaudible} & \textbf{Throughput\tnote{1}} & \textbf{Near} & \textbf{Medium} & \textbf{Far}\\ 
\midrule
R & Lee et al.~\cite{lee2015chirp} & Chirp BOK & 19.5--22 kHz & \yes & 15 bps & \no & \no & \yes \\
A & Digital Voices \cite{lopes2001aerial} & MFSK & 1--3 kHz & \no & 80 bps & \no & \yes & --\tnote{2} \\
R & Nearby \cite{getreuer2018ultrasonic} & DSSS + MFSK & 18.5--20 kHz & \yes & 84 bps & \no & \yes & \yes \\
N & ggwave$_a$ \cite{gerganov2024ggwave} & MFSK & 1.8--6.3 kHz & \no & 268 bps & \no & \yes & --\tnote{2} \\
N & ggwave$_i$ \cite{gerganov2024ggwave} & MFSK & 15--19.5 kHz & \yes & 268 bps & \no & \yes & --\tnote{2} \\
A & HRCSS \cite{cai2022boosting} & OCSS & 18--22 kHz & \yes & 500 bps & \yes & \yes & \yes \\
A & Gon\c{c}alves et al.~\cite{goncalves2017acoustic} & OFDM + QPSK & 0.1--4 kHz & \no & 600 bps & \yes & \no & \no \\
R & PriWhisper \cite{zhang2014priwhisper} & MFSK & 9--17 kHz & \no & 729 bps & \yes & \no & \no \\
\bottomrule
\end{tabular}
  \scriptsize
  \begin{tablenotes}
    \item[1] We list the net data rate, as this is the usable data rate after error correction from the user's perspective.
    \item[2] Maximum distance not specified, therefore no a priori assessment.
  \end{tablenotes}
\end{threeparttable}
\end{table}

\subsection{Summary}
We ultimately obtained a subtotal of six academic implementations for state-of-the art acoustic communication systems.
Additionally, we included two non-academic implementations for comparison from the open-source project ggwave,\footnote{Most-starred ``data-over-sound'' project on GitHub: \url{https://github.com/topics/data-over-sound} (as of 2025-03-03).} in both its audible (\textbf{ggwave$_a$}) and inaudible (\textbf{ggwave$_i$}) configurations.
Thus eight schemes advance to evaluation (\autoref{tab:schemes}).

\section{Evaluation}\label{sec:evaluation}
In this section, we systematically evaluate the generalizability and practical suitability of acoustic data transmission across various use cases. For instance, indoor IoT communication \cite{haus2019enhancing} requires reliable transmission over distances of several meters, whereas mobile payment systems \cite{copsonictechnologies2024copsonic,theamericanbusinessawards2014verifone} operate effectively over just a few centimeters. When integrated into smartphone applications (e.g., Google’s Nearby system \cite{getreuer2018ultrasonic}), acoustic data transmission must accommodate diverse smartphone models with varying audio hardware and frequency responses. Furthermore, transmissions occur under different environmental conditions and must remain robust against various background noises. Our evaluation addresses these considerations systematically.

\subsection{Method}
We conducted four experiments in which we varied the distance (\autoref{sec:experiment-distance}), device models (\autoref{sec:experiment-device-models}), noise interference (\autoref{sec:experiment-noise}), and transmission environment (\autoref{sec:experiment-environment}).
We treated each implementation as a black box according to our system model (\autoref{fig:model}), using it as designed by the authors with their specified parameters.\footnote{Most of the original papers only present one set of PHY parameters. There are two exceptions: (1) PriWhisper \cite{zhang2014priwhisper}, which varied the number of frequencies in an M-FSK scheme—we used the highest number to match their claimed throughput; and (2) DigitalVoices \cite{lopes2001aerial}, which varied PHY parameters solely to assess sound pleasantness, which is outside our scope.}
Our goal was to evaluate each scheme as a complete communication system from a \textit{link layer} (layer two) perspective, including framing, synchronization, and error correction.

We randomly chose the data $b_\mathtt{tx}$ for transmission.
The transmitter converted this data into audio and could utilize the 0-\SI{24}{kHz} baseband freely. Each WAV file was normalized to a specific maximum amplitude before transmitting it from one smartphone to another. 
The received audio was recorded into a WAV file and processed by the receiver implementation, which demodulated the signal to output data $b_\mathtt{rx}$. We then verified whether $b_\mathtt{rx}$ matches $b_\mathtt{tx}$.
This procedure allowed us to treat each scheme as a black box, capable of transmitting data from one device to another. 
Each measurement was repeated at least 20 times for each scheme \cite{jeon2016noncoherent}, while the scheme from Lee et al.~\cite{lee2015chirp} was repeated 100 times due to its smaller transmission duration. We conducted additional measurements in case of measurement errors.
Our experimental setup is shown in \autoref{fig:experimental_setup} and is described in detail in \autoref{sec:meta-appendix} to provide further transparency, including full hardware and software configurations.

\subsubsection{Metrics}
To assess reliability, we introduce the total error rate (TER), a link-layer metric that combines elements of the bit error rate (BER) and the packet error rate (PER), allowing more direct comparisons across different schemes. We also provide the underlying raw BER data in the supplementary material.
Under our system model (\autoref{fig:model}), the TER measures the final error rate of received data bits \emph{after} error correction by comparing $b_\mathtt{rx}$ with $b_\mathtt{tx}$. If the receiver returns an error (e.g., from synchronization failures) instead of data, we treat the transmission as having a 100\% TER.\footnote{Our measurements show that error rates can exceed 50\% when significant portions of the received message are missing. To correctly penalize complete failures---when no message is received---we assign 100\% TER.}
This classification treats such transmissions equal to transmissions where every bit was decoded incorrectly, as both instances represent completely unusable transmissions from the user's perspective.
Previous work on acoustic data transmission often reports only BER \cite{jeon2016noncoherent,satapathy2020silentinformer,goncalves2017acoustic} or PER \cite{zhang2014priwhisper,getreuer2018ultrasonic}, but each has drawbacks.
BER overlooks transmissions that fail to decode, allowing a possible 0\% BER even when most transmissions are undecodable, and PER treats a single bit error the same as a thousand bit errors.
By combining both metrics, TER better reflects real-world reliability, accounting for individual bit errors as well as complete transmission failures.\footnote{In summary, the TER is equal to the BER unless the receiver returns an error instead of data, in which case the TER is 100\%. Illustrative examples (assuming 100 data bits per frame):
\begin{enumerate}[after=\vspace{-0.8\baselineskip}]
    \item Receiving 99 correct bits yields a TER of 1\% (identical to the BER).
    \item Failing to decode entirely yields a TER of 100\% (where the BER is undefined).
    \item Receiving all bits incorrectly also yields a TER of 100\% (identical to the BER).
    \item Across 50 transmissions in case (1) and 50 in case (2), the average TER is 50.5\%, whereas BER is misleadingly low at 1\%, and PER is misleadingly high at 100\%.
\end{enumerate}}

\subsubsection{Parameters}
We conducted four experiments in which we systematically varied the following parameters to determine their impact on reliability. These parameters are the most commonly evaluated ones in prior research using real devices \cite{goncalves2017acoustic,getreuer2018ultrasonic,lee2015chirp,zhang2014priwhisper}. 

\begin{figure}[bp]
\centering
\includegraphics[page=2,width=\columnwidth]{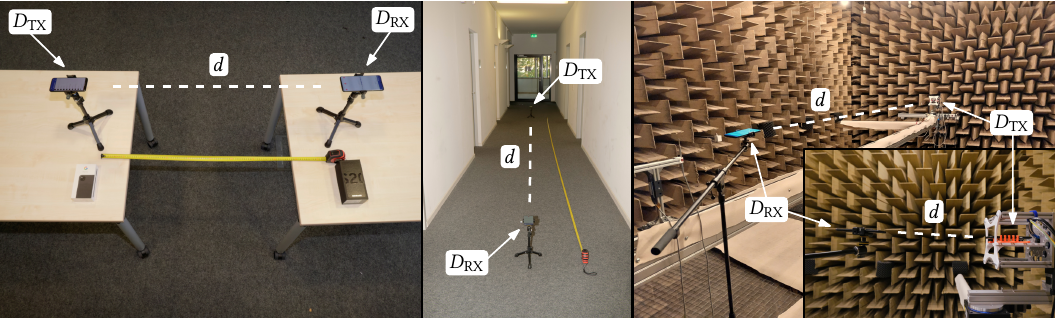}
\caption{\captionheadline{Experimental setup.} The left picture shows our office setup, with device $D_\text{TX}$ transmitting to device $D_\text{RX}$ at distance $d$. The middle picture shows our setup in the office hallway for larger distances from \SI{5}{\meter} to \SI{40}{\meter}. The right pictures show our setup in the anechoic chamber at two different distances.}
\label{fig:experimental_setup}
\Description{Photographs of experimental environments and device placement. The figure is a composite of photos illustrating how the transmitter and receiver devices are positioned in different environments. In the left panel, two small tripods with smartphones face each other across two separate tables. A tape measure runs between the table edges, and a dashed line between the two devices is labeled with the transmission distance $d$. Text labels $D_\text{TX}$ and $D_\text{RX}$ mark the transmitting and receiving devices.

The center panel shows a long, narrow corridor. A transmitter device $D_\text{TX}$ on a tripod is placed near the foreground, and a receiver $D_\text{RX}$ on another tripod is farther down the hallway. A dashed line along the length of the corridor indicates the distance 
$d$ between the two devices.

The right side consists of photos taken in an anechoic chamber with sound-absorbing wedge panels on the walls. One main photo and a smaller inset photos show the transmitter and receiver tripods placed inside the chamber. As in the other panels, labels $D_\text{TX}$, $D_\text{RX}$, and dashed lines labeled $d$ indicate the positions of the devices and the separation distance between them.}
\end{figure}

\begin{itemize}
    \item \textbf{Distance} (\autoref{sec:experiment-distance}). The transmission distance $d$ significantly affects suitability for our designated use case categories. For instance, a scheme failing to transmit reliably over multiple meters is unsuitable for the far-distance use case. We set a default distance of $d=\SI{50}{cm}$, typical for medium-distance use cases.

    \item \textbf{Device models} (\autoref{sec:experiment-device-models}). The transmitter $D_{\text{TX}}$ and receiver $D_{\text{RX}}$ influence transmission reliability, as the audio hardware in smart devices is not designed for data transmission. Each device has a different frequency response, particularly at near-ultrasonic frequencies~\cite{hromadova2022frequency}. We selected the Google Pixel 4a smartphone as the default transmitter $D_{\text{TX}}$ due to its relatively low distortion at a volume setting\footnote{On smartphones, the volume can typically only be set to one of a few discrete volume indices, where the maximum index varies per device.} of $V_{\text{TX}} =$ 19/25, which we experimentally identified to be a good trade-off between high volume and low distortions due to amplifier and speaker non-linearities. The default receiver $D_{\text{RX}}$ was a Samsung Galaxy S20 Ultra smartphone, chosen for its low self-noise.

    \item \textbf{Noise interference} (\autoref{sec:experiment-noise}). Practical transmissions often occur under suboptimal conditions, such as loud ambient noise or interference from how users handle their smartphones. By default, our testing occured in a quiet environment without external noise.

    \item textbf{Environment} (\autoref{sec:experiment-environment}).
The environment influences the multipath profile of the acoustic channel. By default, we evaluated the schemes in a typical indoor setting using quiet office rooms and a hallway for long-distance measurements.
Because these environmental conditions cannot be replicated exactly, we also conducted tests in an anechoic chamber (\autoref{fig:experimental_setup}) without reverberation or ambient noise, offering a controlled and reproducible testing environment.
Finally, we also tested a large lecture room with mechanical ventilation.
The background noise levels averaged \SI{53.60}{dB} SPL in the office, \SI{60.01}{dB} SPL in the lecture room, and \SI{27.65}{dB} SPL in the anechoic chamber (all unweighted RMS).
\end{itemize}

\subsection{Preliminary testing}\label{sec:experiment-basic-success}
\subsubsection{Setup}
Before extensive smartphone-based tests, we performed a preliminary evaluation under near-optimal conditions to confirm each implementation’s functionality.
Using high-quality audio equipment, each scheme was transmitted by a Neumann KH 80 DSP studio speaker and recorded by an Earthworks M23R reference microphone, positioned on-axis at \SI{50}{cm} in an office room (see \autoref{sec:experiment-basic-success-setup} for details).

\subsubsection{Results}
\begin{figure}[hbp]
\centerline{\includegraphics[width=\columnwidth]{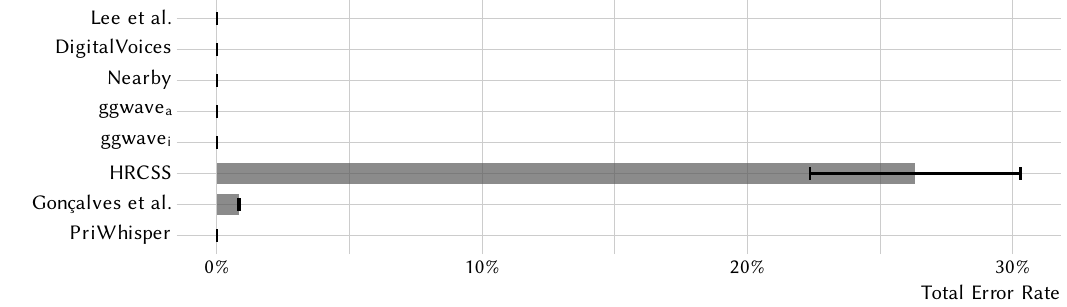}}
\caption{\captionheadline{Preliminary testing in a best-case hardware scenario.} Shorter bars are better. Schemes are sorted by throughput (see \autoref{tab:schemes}).  The error bars show the standard error. Parameters: $d=\SI{50}{cm}$; $D_{\text{TX}}=$ loudspeaker Neumann KH 80 DSP; $D_{\text{RX}}=$ microphone Earthworks M23R; $V_{\text{TX}}=\SI{-40}{dB}$ interface gain; quiet office environment; N=\num{240} measurements.
}
\Description{The figure is a horizontal bar chart comparing total error rates (in percent) for eight acoustic data transmission systems. The x-axis ranges from 0\% to 30\% total error rate. Each system is represented by a separate bar. Only HRCSS and the system by Gon\c{c}alves et al.~had non-zero errors, with HRCSS having most.}
\label{fig:basic-success}
\end{figure}

The scheme by Lee et al., DigitalVoices, Nearby, PriWhisper, and both ggwave variants achieved a perfect 0\% TER  (\autoref{fig:basic-success}).
The scheme from Gon\c{c}alves et al.~was mostly accurate but showed an error rate of 0.8\%, likely due to the scheme being designed for closer distances.
However, the HRCSS implementation produced extremely high TER, which was unexpected given the high-quality equipment.
We unfortunately had to conclude that this implementation is unsuitable for practical tests and therefore excluded it from further testing (see discussion in \autoref{sec:hrcss-discussion}).

\subsection{Impact of different distances}\label{sec:experiment-distance}
\subsubsection{Setup}
Next, we evaluated each scheme over distances from \SI{5}{cm} to \SI{40}{m}. Tests were conducted in both a quiet indoor setting (representative of everyday use) and an anechoic chamber with minimal reverberation and noise. In both environments, we measured distances up to \SI{5}{m}; for distances beyond \SI{5}{m} (up to \SI{40}{m}), we used the hallway adjacent to the office.

\begin{figure}[tbp]
\centering
    \begin{subfigure}[t]{\textwidth}
    \centering
    \includegraphics[width=.95\textwidth]{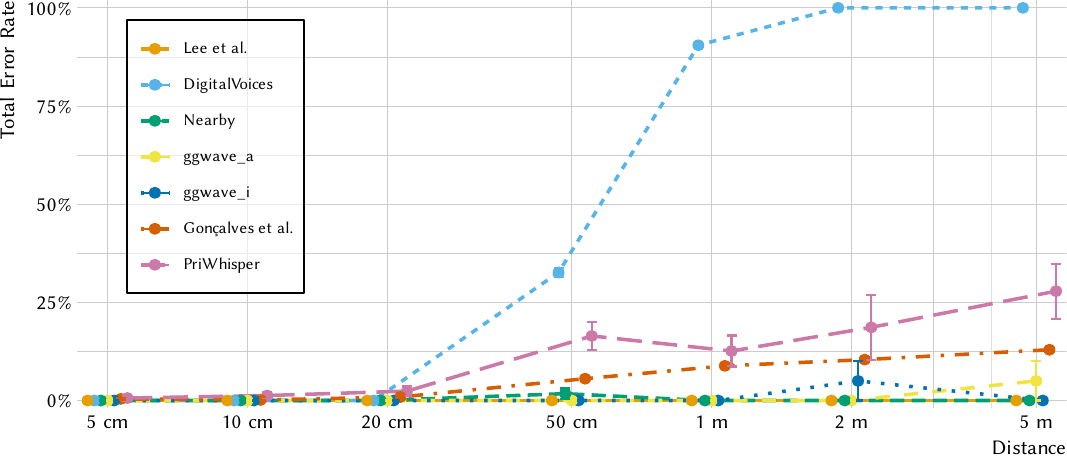}
    \includegraphics[width=.95\textwidth]{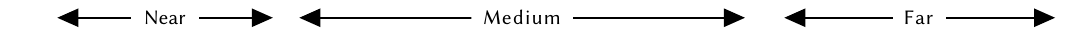}
    \caption{
      Anechoic chamber.
    }
    \Description{Line chart showing total error rate versus distance for seven acoustic data transmission systems in an anechoic chamber. The x-axis spans from a few centimeters to several meters, and the y-axis shows the Total Error Rate from 0\% to 100\%. All systems exhibit very low error at short distances; as distance increases, error rates for Digital Voices rise sharply toward 100\%, while others maintain comparatively lower errors over a larger range of distances.}
    \label{fig:distance-anechoic}
    \end{subfigure}
    \begin{subfigure}[b]{\textwidth}
        \centering
        \vspace{0.9cm}
        \includegraphics[width=.95\textwidth]{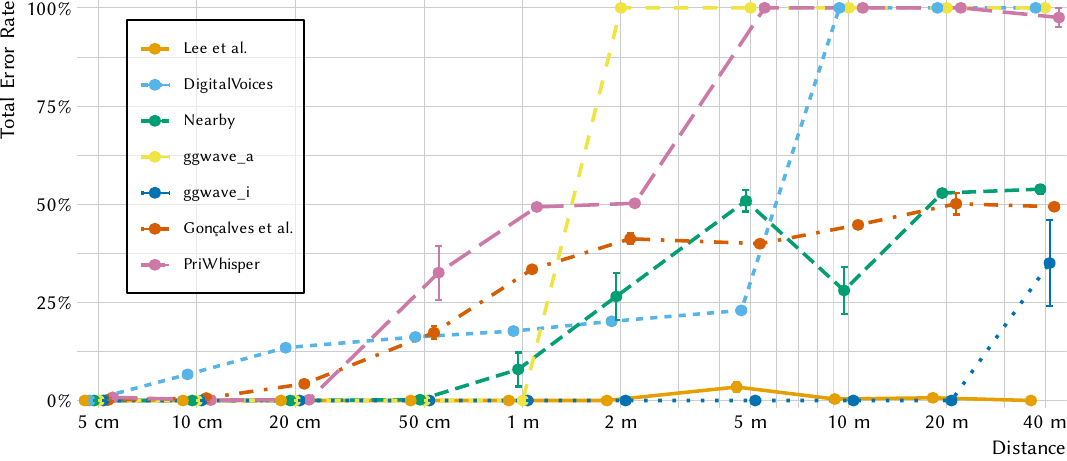}
        \includegraphics[width=.95\textwidth]{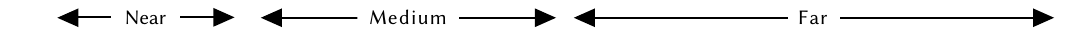}
        \caption{
          Office environment.
        }
        \Description{Line chart showing total error rate versus distance for seven acoustic data transmission systems in an office environment. The x-axis spans from a few centimeters to several meters, and the y-axis shows the Total Error Rate from 0\% to 100\%. All systems work with almost no errors at very short range, but as distance increases, most systems’ error rates rise substantially.}
        \label{fig:distance-office}
    \end{subfigure}

\caption{\captionheadline{Reliability of acoustic transmissions with varying distances.} Smaller TER is better. The anechoic chamber allows for measurements up to a maximum distance of \SI{5}{m}. In the office environment, we were able to measure up to a distance of \SI{40}{m}, but the measurements at \SIrange{10}{40}{m} were conducted in the hallway. The distance axis is logarithmic, but we only measured at the indicated discrete distances. The error bars show the standard error. The data points of different schemes are dodged to the side to avoid overlap. Parameters: $D_{\text{TX}}=$ smartphone Pixel 4a; $D_{\text{RX}}=$ smartphone Samsung S20 Ultra; $V_{\text{TX}}=$ volume index 19/25; anechoic chamber (\autoref{fig:distance-anechoic}) and quiet office environment (\autoref{fig:distance-office}); N=\SI{3740} measurements.}
\Description{A two-part image with a line plot each showing the Total Error Rate in the anechoic chamber and in an office environment, for varying distances.}
\label{fig:distance-results}
\end{figure}

\subsubsection{Results}
In the \textit{anechoic chamber} (\autoref{fig:distance-anechoic}), most schemes achieved low error rates at short distances.
The scheme from Lee et al.~maintained a zero error rate across all tested distances.
Both ggwave schemes performed nearly perfectly, with slight exceptions at \SI{2}{m} and \SI{5}{m}.
Nearby also excelled, aside from a small outlier at \SI{50}{cm}.
Gon\c{c}alves et al.'s scheme kept error rates under 0.5\% for distances below \SI{10}{cm}, but rose to around 13\% at longer distances.
DigitalVoices and PriWhisper stayed below 1\% at close range, then deteriorated notably beyond \SI{20}{cm}.

In the \textit{office environment} (\autoref{fig:distance-office}), error rates increased noticeably due to stronger multipath effects and ambient noise.
DigitalVoices, PriWhisper, Nearby, and Gon\c{c}alves et al.'s scheme all showed a strong distance dependency, degrading over longer ranges.
At short distances up to \SI{10}{cm}, however, all schemes maintained a TER under 0.5\% except DigitalVoices.
Although DigitalVoices performed worse at close range than in the anechoic chamber, it surprisingly improved between \SIrange{50}{500}{cm}, showing that the DigitalVoices receiver is not yet fully optimized. We investigated the DigitalVoices discrepancy and observed that post-processing and amplifying the recordings significantly reduced its error rates in the anechoic environment. However, we did not include post-processed results in our plots to compare the schemes as described in their publications.
Nearby performed well within \SI{1}{m}, but became unreliable at larger distances, including an outlier at \SI{10}{m}, likely due to the transition into the hallway.
This non-monotonic distance behavior is consistent with Getreuer et al.'s original Nearby evaluation \cite[Fig.~13]{getreuer2018ultrasonic}, which attributes such variability to nontrivial multipath behavior.
Lee et al.'s scheme performed exceptionally across all distances, with only a minor outlier at \SI{5}{m}.
The inaudible ggwave variant remained effective up to \SI{20}{m}, whereas the audible variant only worked up to \SI{1}{m} before failing to decode.

\subsection{Impact of different device models}\label{sec:experiment-device-models}
\subsubsection{Setup}
This experiment assessed how well acoustic data transmission schemes perform across a broad set of smartphones in a typical indoor environment.
We selected five smartphones representing a broad spectrum of release dates, prices, and manufacturers (\autoref{tab:device-models-smartphones}).
We tested each scheme using 21 different transmitter-receiver combinations.
Our setup involved one smartphone as the transmitter, positioned \SI{50}{cm} away on-axis from two to three simultaneous receivers to make the measurement process more efficient.
As we had two Pixel 4a smartphones, we also evaluated performance with identical transmitter and receiver models.

\begin{table}[bt]
\centering
\caption{\captionheadline{Smartphones in the device models experiment} in \autoref{sec:experiment-device-models}. The volume column indicates the selected and maximum volume indices for each device. We set each device to a volume of ca.~75\% of its maximum volume, which we experimentally identified to be a good trade-off between achieving high volume while maintaining low distortions due to amplifier and speaker non-linearities.}
\small
\begin{tabular}{lllll}
\toprule
\textbf{Device Name} & \textbf{Origin} & \textbf{Release Date} & \textbf{Price} & \textbf{Volume}\\ 
\midrule
Google Pixel 4a & USA & Oct 2020 & 349€ & 19/25\\
Google Pixel 6 Pro & USA & Oct 2021 & 900€ & 19/25\\
Huawei Nexus 6P & China & Sep 2015 & 700€ & 11/15\\
Oppo Reno 6 & China & Sep 2021 & 499€ & 12/16\\
Samsung Galaxy S20 Ultra & South Korea & Feb 2020 & 1300€ & 11/15\\
\bottomrule
\end{tabular}
\label{tab:device-models-smartphones}
\end{table}

\subsubsection{Results}
\begin{figure}[bp]
\centerline{\includegraphics[width=1\columnwidth]{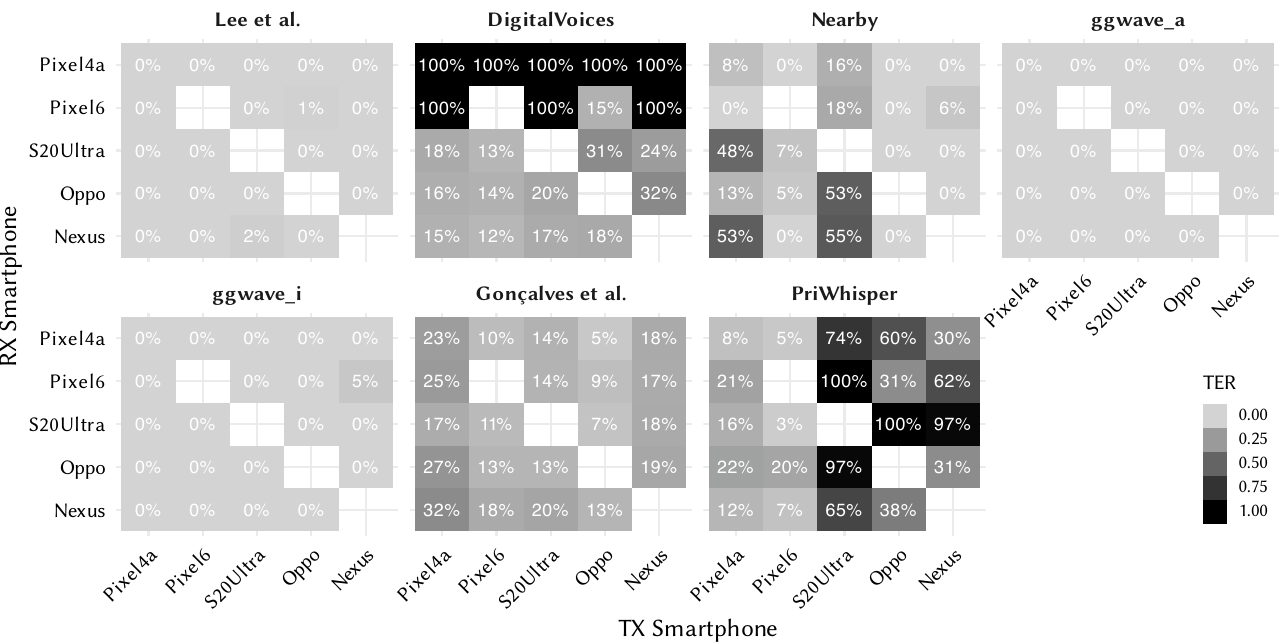}}
\caption{\captionheadline{Reliability of acoustic transmissions for different device combinations.} Smaller TER is better. Schemes are sorted by throughput. The heatmap shows each scheme's reliability for 21 different TX-RX combinations. Parameters: $d=\SI{50}{cm}$; $D_{\text{TX}}$/$D_{\text{RX}}$/$V_{\text{TX}}=$ see \autoref{tab:device-models-smartphones}; office environment; N=\SI{4620} measurements.} 
\Description{Heatmap matrix of Total Error Rates for seven acoustic schemes across all pairs of five Android smartphones used as transmitter and receiver. Darker cells indicate higher error and lighter cells lower error. Some schemes show consistently low errors for most device combinations, while others fail on many pairs or only work reliably for a few specific transmitter–receiver combinations.}
\label{fig:heatmap-models}
\end{figure}

As shown in \autoref{fig:heatmap-models}, Lee et al.'s scheme and both ggwave variants consistently exhibited very low error rates across all device combinations.
DigitalVoices showed varying performances: the best results occurred with the Pixel 6 Pro transmitting to the Nexus 6P (TER $\approx$ 12\%), while the Pixel 4a and the Pixel 6 Pro largely failed as receivers.
Nearby generally performed well, except when using the Samsung S20 as the transmitter and for some transmissions of the Pixel 4a.
PriWhisper was effective in certain combinations, notably with the Pixel 6 Pro as the transmitter (TER as low as 3\%), but failed in others due to synchronization issues.
Its poor overall performance can be attributed to the \SI{50}{cm} device distance, which highlights its limitations outside its intended usage range of a few cm. In this case, the impact of different device models on the reliability varied significantly, ranging from satisfactory to completely ineffective.
The performance of Gon\c{c}alves et al.'s scheme was more consistent, with error rates from 5\% (best case: Oppo Reno 6 to Pixel 4a) to 32\% (worst case: Pixel 4a to Nexus 6P).

Overall, the Pixel 6 Pro as the transmitter consistently resulted in the lowest error rates. The Samsung Galaxy S20 Ultra generally performed well as a receiver, except with certain transmitters for DigitalVoices and PriWhisper. This highlights a large diversity in device characteristics, which complicates the generalization of some schemes.

\subsection{Impact of different types of noise and device handling}\label{sec:experiment-noise}
\subsubsection{Setup}

The preceding tests were conducted in relatively quiet environments, but practical scenarios often contain different types of interference, which can degrade the reliability of acoustic transmissions either due to the ambiance and the surroundings, or due to human handling of the devices.
In this experiment, we aimed to assess the influence of different kinds of noises and handling on the receiver.
We used the Pixel 4a as the transmitter and the Samsung Galaxy S20 Ultra as the receiver, at a distance of  \SI{50}{cm} aligned on-axis in a quiet indoor environment.
We performed experiments for each of the following seven types of noise and handling, designed to mimic real-world disturbances:

\begin{itemize}
    \item \textbf{Ambient.} Playing ambient field recordings from a café, a train station, and a market place on a Neumann loudspeaker \SI{1}{m} away from the receiver. The experiment was conducted separately for all three ambient sounds (see \autoref{sec:experiment-noise-setup} for detailed setup information).
    \item \textbf{Circle movement.} Shuffling the receiver in a circular motion on the table, which varies the distance by $\pm$\SI{5}{cm}.
    \item \textbf{Clapping.} Clapping hands near the receiver every \SI{2}{s}.
    \item \textbf{Pocket.} Inserting the receiver into and removing it from a trouser pocket every \SI{2}{s}, aligned so that the transmitter's speaker faces the receiver in the pocket.
    \item \textbf{Up and down.} Lifting and firmly placing the receiver on the table approximately every \SI{2}{s}.
    \item \textbf{Opposite.} The smartphones are positioned with their microphones and speakers facing opposite directions. This arrangement simulates two people holding phones while facing each other.
    \item \textbf{Side-by-side.} The smartphones are placed next to each other, with their long edges adjacent.
    \item \textbf{Handheld.} The transmitter smartphone is held by a person, introducing the natural movement of typical handheld use.
\end{itemize}

\subsubsection{Results}
\begin{figure}[btp]
  \centerline{\includegraphics[width=1\columnwidth]{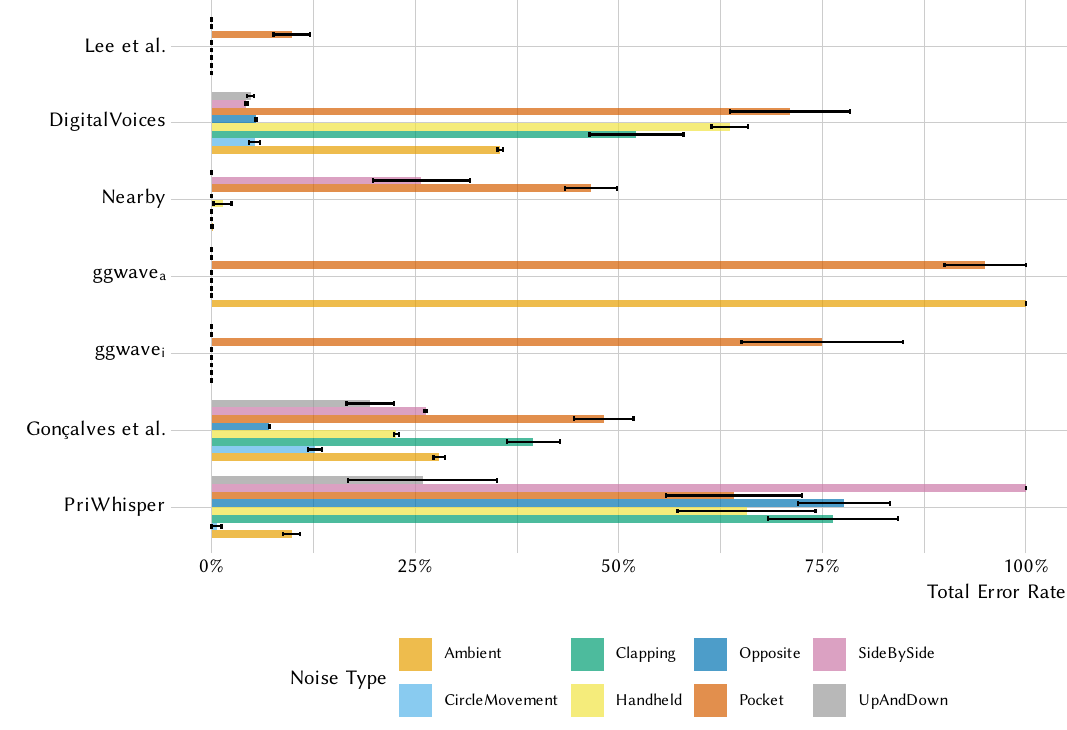}}
  \caption{\captionheadline{Reliability for different types of noise disturbing the transmission.} Smaller bars are better. Schemes are sorted by throughput. For the ambient noise, we performed three separate experiments with different field recordings. The error bars show the standard error. Parameters: $d=\SI{50}{cm}$; $D_{\text{TX}}=$ smartphone Pixel 4a; $D_{\text{RX}}=$ smartphone Samsung S20 Ultra; $V_{\text{TX}}=$ volume index 19/25; office environment with different types of controlled noise (see \autoref{sec:experiment-noise}); N=\SI{2200} measurements.}
  \Description{Grouped horizontal bar chart comparing Total Error Rates of seven acoustic transmission schemes under different types of disturbing noise.}
  \label{fig:noise-results}
  \end{figure}

All schemes experienced significant performance degradation when the smartphone was pocketed during transmission (\autoref{fig:noise-results}).
However, Nearby, the inaudible ggwave variant, and Lee et al.'s scheme were robust against all other noise types tested.
The audible ggwave variant was affected by ambient noise, which made it completely unusable, likely because the ambient noise interfered with the transmission bandwidth.
Clapping dramatically impaired DigitalVoices, PriWhisper, and Gon\c{c}alves et al.'s scheme, rendering them largely ineffective. Noise interactions with the table, whether through circular movement or repetitive lifting and placing, had a relatively minor impact on performance. Notably, DigitalVoices struggled significantly with clapping and pocketing noises (TER > 50\%), yet coped better with table-related movements (TER $\approx$ 6\%).
The side-by-side orientation was more detrimental to reliability than the opposite orientation, although not all schemes were affected.

  \begin{figure}[btp]
  \centerline{\includegraphics[width=1\columnwidth]{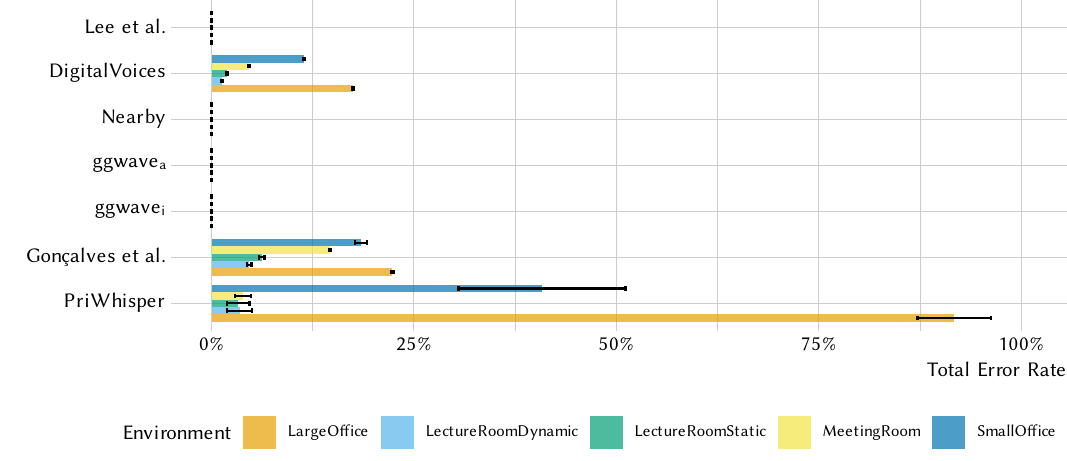}}
  \caption{\captionheadline{Reliability in different transmission environments.} Smaller bars are better. Schemes are sorted by throughput. The error bars show the standard error. Parameters: $d=\SI{50}{cm}$; $D_{\text{TX}}=$ smartphone Pixel 4a; $D_{\text{RX}}=$ smartphone Samsung S20 Ultra; $V_{\text{TX}}=$ volume index 19/25; N=\SI{1100} measurements.}
  \Description{Grouped horizontal bar chart comparing Total Error Rates of seven acoustic transmission schemes in different transmission environments.}
  \label{fig:environment-results}
  \end{figure}

\subsection{Impact of different environments}\label{sec:experiment-environment}
\subsubsection{Setup}

To determine the influence of the environment and its multipath profile, we measured transmissions between a Pixel 4a (transmitter) and a Samsung Galaxy S20 Ultra (receiver) aligned on-axis at \SI{50}{cm}. The evaluation took place in several locations: small and large offices, a meeting room, and a lecture room. In the lecture room, we evaluated both a static setup and a dynamic one in which people walked quietly throughout the room while maintaining an unobstructed line of sight between the devices, dynamically changing the multipath environment (see \autoref{sec:experiment-environment-setup}).

\subsubsection{Results}
As shown in \autoref{fig:environment-results}, Lee et al.'s scheme, Nearby, and both ggwave variants achieved perfect reliability in all tested environments. The remaining schemes performed worst in the offices, but showed improvement in the meeting and lecture rooms. Interestingly, quiet movements in the vicinity do not seem to negatively impact reliability.

\subsection{Summary for each use case}

\begin{figure}[bp]
\centerline{\includegraphics[width=1\columnwidth]{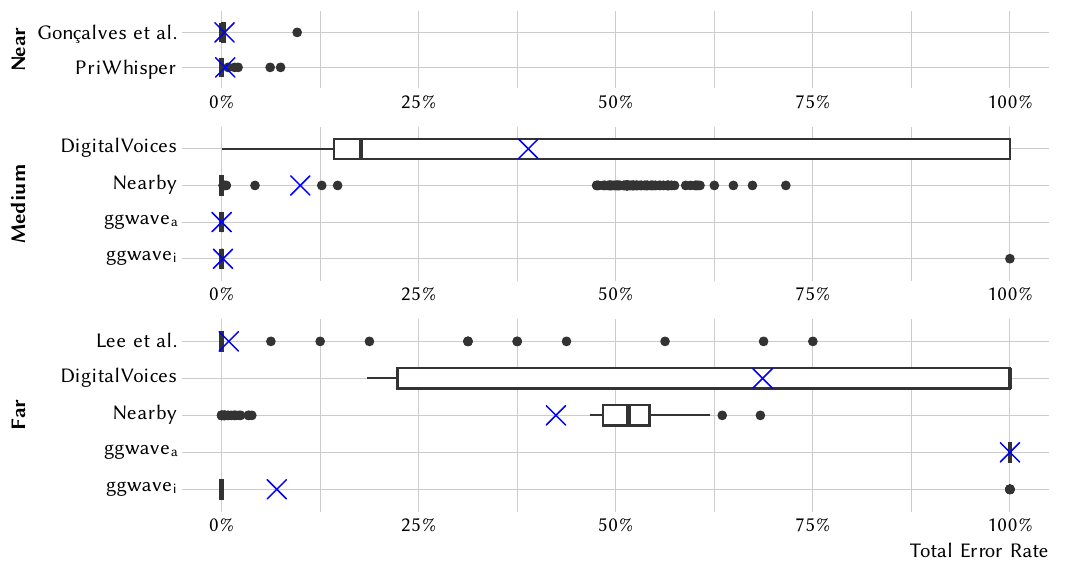}}
\caption{\captionheadline{Summary of our measurement results from a quiet indoor environment} (\autoref{sec:experiment-distance}, \autoref{sec:experiment-device-models}, and \autoref{sec:experiment-environment}), for the three use case categories (defined in \autoref{sec:use-cases}  and mapped in \autoref{tab:schemes}). Smaller TER is better. Schemes are sorted by throughput in each category. The box plots display the median (black bar), mean (blue cross), and outliers (black circles), with hinges marking to the 25th and 75th percentiles. Parameters: $d=$ variable (according to the scenarios in \autoref{sec:use-cases}); $D_{\text{TX}}$/$D_{\text{RX}}$/$V_{\text{TX}}=$ variable according to \autoref{tab:device-models-smartphones}; quiet office environment; N=\SI{3300} measurements.
}
\Description{Grouped horizontal box plots showing the distribution of Total Error Rate for several acoustic transmission schemes in a quiet office environment, separated into three use case categories: Near, Medium, and Far. Within each category, different schemes have box plots spanning 0–100\% error, illustrating that some methods remain reliably low-error at close range, while error rates generally increase and become more variable for medium and far scenarios, with several schemes approaching very high error for distant use.}
\label{fig:use-cases-results}
\end{figure}

We present a summary of our previous measurement results by scheme and use case in \autoref{fig:use-cases-results}. This plot shows how each scheme performs for its intended use case (as per \autoref{tab:schemes}) in typical indoor environments, therefore we excluded measurements in the anechoic chamber and with the high-quality studio equipment, as they do not represent typical usage. We discuss these results in more detail in \autoref{sec:rq2-discussion}.

\subsubsection{Outliers}
For the inaudible ggwave variant, one transmission failed during the medium-distance use case ($d=\SI{50}{cm}$, $D_{TX}$ = Nexus 6P, $D_{RX}$ = Pixel 6 Pro), and some transmissions failed for the far-distance use case ($d=\SI{40}{m}$, $D_{TX}$ = Pixel 4a, $D_{RX}$ = Samsung S20 Ultra). 
The outliers for the scheme by Lee et al.~were mostly at a distance of \SI{5}{m} as shown in \autoref{fig:distance-office}.
Nearby performed well in the medium-distance use case (TER median = 0\%), except for a few anomalies in specific device combinations (e.g., $D_{TX}$ = Samsung S20 Ultra, $D_{RX}$ = Nexus 6P).

\section{Discussion}\label{sec:discussion}

\subsection{Obtaining software implementations is challenging}\label{sec:discussion-obtaining}
To answer our first research question (\textbf{RQ1}) of how researchers can obtain working implementations of acoustic data transmissions systems for evaluation, we conducted a systematic literature study and documented our experience in trying different acquisition methods (\autoref{sec:obtaining-implementations}).  
Out of the 31 publications identified in our survey, none referenced a publicly available implementation, and only four authors provided their implementation upon request.
This lack of accessible research artifacts severely hinders
replication of research in the field of acoustic data transmission, thereby slowing scientific progress.
Without straightforward access to verify and build on existing work, demonstrating improvements over the state of the art becomes significantly more challenging.

We also attempted to re-implement the schemes based on the description in the paper, a process that is both time-consuming and rarely undertaken by other researchers. 
During this effort, we encountered several inaccuracies and omissions in the original publications, which often required educated guesses to resolve (\autoref{sec:reimplementation-challenges}).
This challenge is not unique to our field but has been documented in other areas of computer science as well~\cite{collberg2016repeatability}.
Such re-implementations carry a high risk of errors or deviations from the original, which is frustrating given the significant time investment.
To improve replicability and transparency in research, we encourage other researchers to follow our suggestions in \autoref{sec:discussion-suggestions} and publish their software artifacts.

\subsection{Generalizability towards ad-hoc smartphone communication use cases}\label{sec:rq2-discussion}

To assess the practicality of the acoustic data transmission schemes proposed in previous studies (\textbf{RQ2}), we evaluated how well these schemes generalize towards the three use case categories defined in \autoref{sec:use-cases}.
For each use case category, we first reviewed the original publications of each scheme to identify their a priori suitability based on their specified throughput and maximum distance.
For example, the scheme by Lee et al. only has a throughput of 15 bps, so by design it is not suitable for transmission of public keys or small files at near distances, as this would take far too long.
In many cases, our experiments revealed limited practical suitability of the schemes, which was surprising given the claims in their original publications.
\autoref{tab:scheme-results} shows our a posteriori assessment of each scheme based on our evaluation.
In particular, all highlighted results in the table represent novel insights which were previously not known based on the limited testing in the current literature. This discrepancy shows the importance of extensive testing across a range of distances and devices to demonstrate practical suitability.

For \textit{far distances} (up to \SI{40}{\meter}), the only reliable schemes were the inaudible ggwave variant, effective up to \SI{20}{\meter}, and the scheme by Lee et al., which maintained reliability up to \SI{40}{\meter}.
All other schemes failed at these distances.
At \textit{medium distances}, both ggwave variants were effective, with the inaudible one being particularly robust.
The Nearby scheme also performed well but struggled to generalize over all tested device combinations.
The other schemes are not suited for this use case, either because of high error rates or because their throughput is not high enough.
For the \textit{near-distance} use case, only PriWhisper and the scheme by Gon\c{c}alves et al.~had sufficient throughput, while maintaining low error rates. However, they did experience some erroneous packets, indicating a need for better or additional error correction techniques.

\begin{table}[bt]
\centering
\caption{\captionheadline{Suitability of each scheme for different use case categories (RQ2).} Schemes are sorted by throughput. A scheme is considered suitable (\yes) if its mean TER is approximately 0\%. Otherwise, it is unsuitable (\no). When only a few outliers appear and the median TER remains at 0\%, the scheme may still be suitable but requires additional modifications (e.g., error correction). The highlighted entries indicate novel results that changed compared to prior knowledge (\autoref{tab:schemes}). The first column indicates how we obtained the implementation: P -- referenced in the paper; A -- by contacting the authors; R -- by re-implementing them ourselves; N -- non-academic solution for comparison.}
\label{tab:scheme-results}
\small
\begin{threeparttable}
\begin{tabular}{@{}c l@{\hspace{0.1em}}r@{\hspace{0.7em}}r@{\hspace{0.7em}}c@{\hspace{0.5em}}r@{\hspace{0.5em}}|c@{\hspace{0.7em}}c@{\hspace{0.7em}}c}\\
\toprule
\textbf{Via} & \textbf{Name} & \textbf{Frequencies} & \textbf{Bandwidth} & \textbf{Inaudible}  &\textbf{Throughput\tnote{1}} & \textbf{Near} & \textbf{Medium} & \textbf{Far}\\ 
\toprule
R & Lee et al.~\cite{lee2015chirp} & 19.5--22 kHz & 2.5 kHz & \yes & 15 bps & \no & \no & \yes\\
A & Digital Voices \cite{lopes2001aerial} & 1--3 kHz & 2 kHz & \no & 80 bps & \no & \cellcolor{blue!30}\no  & \cellcolor{blue!30}\no\\
R & Nearby \cite{getreuer2018ultrasonic} & 18.5--20 kHz & 1.5 kHz & \yes & 84 bps & \no & \yes\tnote{2} & \cellcolor{blue!30}\no\\
N & ggwave$_a$ \cite{gerganov2024ggwave} & 1.8--6.3 kHz & 4.5 kHz & \no & 268 bps & \no & \yes & \cellcolor{blue!30}\no\\
N & ggwave$_i$ \cite{gerganov2024ggwave} & 15--19.5 kHz & 4.5 kHz & \yes & 268 bps & \no & \yes & \cellcolor{blue!30}\yes\\
A & HRCSS \cite{cai2022boosting} & 18--22 kHz & 4 kHz & \yes & 500 bps & \cellcolor{blue!30}\no & \cellcolor{blue!30}\no & \cellcolor{blue!30}\no\\
A & Gon\c{c}alves et al.~\cite{goncalves2017acoustic} & 0.1--4 kHz & 3.9 kHz & \no & 600 bps & \yes\tnote{2} & \no & \no\\
R & PriWhisper \cite{zhang2014priwhisper} & 9--17 kHz & 8 kHz & \no & 729 bps & \yes\tnote{2} & \no & \no\\
\bottomrule
\end{tabular}
  \scriptsize
  \begin{tablenotes}
    \item[1] We list the net data rate, as this is the usable data rate after error correction from the user's perspective.
    \item[2] Requires further error correction.
  \end{tablenotes}
\end{threeparttable}
\end{table}

None of our analyzed schemes are suitable for all use cases (see \autoref{tab:scheme-results}), as different applications prioritize different trade-offs between reliability, throughput, latency, and audibility. Some schemes offer flexibility; for instance, ggwave can be configured for higher throughput at the cost of reliability or shifted to the near-ultrasonic range to achieve inaudibility.

For applications like secure pairing or mobile payments, where data integrity is paramount, a low TER is the most critical requirement, while high throughput is secondary. The scheme from Lee et al.~and PriWhisper are well-suited for this context, although PriWhisper may benefit from additional error correction to guarantee performance in challenging environments.
In contrast, use cases such as multimedia applications and content delivery (e.g., second-screen services, location-based information) often prioritize high throughput and inaudibility to ensure a smooth user experience, and can typically tolerate minor, correctable errors. The inaudible ggwave variant is a candidate here, though it cannot reach data rates higher than \SI{268}{bps}. Notably, schemes designed for higher data rates generally struggled to perform reliably across varied indoor environments.

Given that high throughput generally decreases reliability, a robust design pattern for many practical applications is to treat the acoustic channel as a low-bandwidth trigger: a short, reliable acoustic message (e.g., a unique identifier or hash) can initiate a connection over a high-throughput channel like Wi-Fi or cellular to transfer the bulk of the data.

\subsection{Practical challenges for acoustic data transmission}\label{sec:practical-challenges}

Our measurements show that acoustic data transmission on commodity smartphones and IoT devices face challenges that rarely appear in simulations or controlled lab settings (\textbf{RQ3}).  
Because audio hardware and propagation channels differ widely across devices and environments, there is no single transmission specification.  
Rather than prescribing a new modem architecture, we therefore summarize the \emph{design requirements} that any future scheme should respect for real-world applicability.

\subsubsection{Severe multipath propagation}
\label{sec:severe-multipath}

Reverberation produces long delay spreads due to multipath propagation at the relatively slow speed of \SI{343}{m/s}.
Because this delay spread depends on the environment and geometry, a scheme that performs well in one setting may behave differently in another.
We observed that even in the same environment, device position and orientation can significantly affect reliability, leading to considerable error rate variability. These findings support the observation by Getreuer et al.~that acoustic transmission reliability is sensitive to environmental conditions~\cite[section 8A]{getreuer2018ultrasonic}.
\textbf{Design requirement:} Operate correctly with delay spreads up to tens of milliseconds \cite{shi2023longrange,lee2015chirp} and the resulting inter-symbol interference (e.g., using guard intervals).

\subsubsection{Device heterogeneity}
Speaker and microphone responses, amplifier gains, and nonlinearities vary widely: our tests reveal up to 30 dB SPL spread between phones at the same volume setting.
Prior research often considered only a specific device model, limiting its practical generalizability.
Most acoustic links are unidirectional, so no feedback-based gain control is possible.  
Developers can only adjust volume relative to each device's maximum output, and our tests show that most devices suffer from nonlinear distortions at volume settings above approximately 75\%, though exact thresholds vary.
\textbf{Design requirement:} Withstand at least 30 dB SNR fluctuation and moderate non-linear distortion without closed-loop calibration (e.g., using normalization and robust modulation).

\subsubsection{Limited bandwidth}
IoT audio hardware typically employs ADCs and DACs with sampling rates of \SI{44.1}{kHz} or \SI{48}{kHz}, which limits the practical bandwidth to less than \SI{22}{kHz}. When transmissions must be inaudible, the usable bandwidth reduces to less than \SI{4}{kHz}, and audio hardware is highly frequency selective in this band \cite{hromadova2022frequency}.
\textbf{Design requirement:} When choosing the near-ultrasonic band for inaudibility, prepare for strong frequency-selectivity and limited throughput.

\subsubsection{Ambient noise}
In contrast to RF communication, the acoustic frequency band is exposed to a wide range of common noise sources---such as traffic, machinery, human speech, and noises from device handling---that occupy a significant share of the available spectrum.
Both burst and stationary noise components are common.
\textbf{Design requirement:} Maintain link integrity in the presence of continuous and impulsive noise (e.g., using strong error correction and time interleaving).

\subsubsection{Throughput vs reliability}

Some of our surveyed publications in \autoref{tab:survey} pursue very high throughput, but our evaluation shows that increased throughput typically comes at the expense of higher error rates. 
Commercial systems often target $\le$\SI{200}{bps} \cite{cueaudioinc.2024cue,getreuer2018ultrasonic,sonos2022near,stimshop2024wius}, likely because of reliability concerns in practice.
\textbf{Design requirement:} Favor robustness and aggressive error control unless bi-directional link adaptation is feasible (e.g., HRCSS-style probing), which is often impossible in one-way IoT scenarios.

\subsection{Comparison with original evaluations} Several schemes we tested were also evaluated by their original authors.
Although our experimental setup differs intentionally to assess generalizability, we compare our results to determine whether the original findings are \textit{conceptually} replicable \cite{shrout2018psychology}.
Overall, our results align with the original evaluations---considering that deviations are expected due to differences in device characteristics, environments, and our re-implementations.
The only exception is HRCSS \cite{cai2022boosting}, whose provided implementation yielded impractically high error rates even under optimal conditions.
DigitalVoices and ggwave lack prior reliability evaluations.

\subsubsection{PriWhisper}
Zhang et al.~\cite{zhang2014priwhisper} conducted a basic reliability evaluation indoors using a Samsung Galaxy S3 and a Google Nexus S, reporting a packet error rate of 0.5\%.
In our near-distance tests, we observed a similar bit error rate (around 0.5\%), though our packet error rate was somewhat higher.
Despite these differences, our results are comparable, given that our re-implementation might not be as optimized as the original and that we used different smartphone models.

\subsubsection{Gon\c{c}alves et al.}
Gon\c{c}alves et al.~\cite{goncalves2017acoustic} evaluated their scheme indoors on two portable computers, reporting bit error rates of approximately 3\% at \SI{10}{\centi\meter} and 25\% at \SI{100}{\centi\meter}.
Our experiments revealed a similar trend.
Although their noise tests (using background music from a nearby loudspeaker) lack detailed parameters, our noise experiments at \SI{50}{\centi\meter} confirmed a comparable degradation in reliability.

\subsubsection{HRCSS}\label{sec:hrcss-discussion}
Cai et al.'s HRCSS \cite{cai2022boosting} initially appeared promising, but the provided MATLAB implementation failed our preliminary tests with high-quality equipment.
To rigorously justify its exclusion, we analyzed the provided MATLAB source code in detail. We found that the high error rates stemmed from synchronization failures. In particular, the preamble detection algorithm in the implementation deviates from the method described in the HRCSS paper \cite{cai2022boosting}.
Specifically, errors in the code cause the receiver to incorrectly calculate and select signal peaks, preventing it from correctly identifying the start of a transmission. Although we did not attempt to fix the code, we verified synchronization as the root cause by manually supplying the correct preamble start index to the decoder, which substantially improved the TER.

We note that HRCSS occasionally produced excellent results, although its performance was highly inconsistent across distances, volumes, and devices.
While their paper mentions a Java implementation used in their evaluations, we only received the MATLAB code, so we were unable to test the Java version.
This is particularly unfortunate because their study uniquely benchmarks their scheme against an existing one (by Lee et al.~\cite{lee2015chirp}) using real devices.

\subsubsection{Lee et al.} Lee et al.~\cite{lee2015chirp} evaluated their scheme on Android devices in indoor far-distance scenarios, achieving packet success rates above 97\% at distances of \SI{10}{\meter} and \SI{25}{\meter}. Our experiments confirmed these high success rates (over 98\% at \SI{10}{\meter}, \SI{20}{\meter}, and \SI{40}{\meter}). Additionally, our compatibility tests across 21 device combinations---using smartphones as transmitters at \SI{50}{\centi\meter}---demonstrated excellent robustness and versatility.

\subsubsection{Nearby} Getreuer et al.~\cite{getreuer2018ultrasonic} tested Nearby with a Nexus 6 transmitter and Nexus 5/5X receivers, reporting packet success rates over 94\% up to \SI{2}{\meter}, though error rates increased at larger distances. Our tests confirmed low error rates below \SI{1}{\meter} and higher errors at \SI{1}{\meter} and \SI{2}{\meter}, likely due to differences in our re-implementation, device models, and packet sizes.
Another difference is that Getreuer et al. employed a form of error correction by repeating each transmission three times and using the redundancy to correct bit errors, but we did not evaluate this as it effectively reduces throughput to a third. The packet error rate, however, improves when using this form of error correction. 

\subsection{Towards more rigorous and reproducible evaluation}\label{sec:discussion-suggestions} 

Our evaluation reveals that many existing acoustic data transmission systems struggle in practical scenarios.
This highlights a critical need for more rigorous and reproducible evaluation methodologies to properly assess how new schemes generalize to real-world conditions. The challenges in replicating prior work---due to environmental variations and incomplete documentation---have historically limited comparative analysis, with most studies evaluating new schemes in isolation.

To address this, we present a set of guidelines for more rigorous and reproducible evaluation. The black-box analysis framework we detail in \autoref{sec:evaluation} serves as a practical implementation of these principles. A key advantage of our methodology is its generality; being system-agnostic, it can be applied to any unidirectional acoustic data transmission system, ensuring its relevance for evaluating future schemes.
We encourage future publications to adopt this framework.

\subsubsection{Prioritize real-world evaluation over simulation.} While simulations are useful for initial design, final evaluations must be conducted on real devices in scenarios representative of the target use case. Our approach in \autoref{sec:evaluation} provides a template for how to systematically test the impact of key practical challenges:

\begin{itemize}
    \item \textbf{Distance.} Test at various distances relevant to the intended application.
    \item \textbf{Device heterogeneity.} Use a diverse set of transmitter and receiver devices (e.g., different smartphone models or IoT hardware).
    \item \textbf{Environmental noise.} Assess performance against both ambient noise (using field recordings or white noise) and intermittent burst noise (e.g., from device handling).
    \item \textbf{Multipath effects.} Evaluate in environments of varying sizes and acoustic properties.
\end{itemize}

\subsubsection{Focus on comprehensive and relevant metrics.} The TER is arguably the most critical metric, as it directly reflects the system's reliability from the perspective of a developer wanting to integrate acoustic data transmission schemes into their IoT or smartphone projects. Other factors such as audibility and latency do not have to be experimentally determined as they follow directly from the system's design. In time-critical applications, the additional processing latency during encoding and decoding can also be of importance.

\subsubsection{Ensure replicability through open science practices.} To enable verification and build upon prior work, we advocate for the following best practices \cite{bajpai2019dagstuhl,collberg2016repeatability,nosek2015promoting,salehzadehniksirat2023changes,shrout2018psychology}:

\begin{itemize}

\item \textbf{Publish software artifacts.}
Release complete implementations (transmitter and receiver code), ideally along with clear run instructions and analysis scripts. None of the papers we reviewed provided their code, forcing error-prone and time-consuming re-implementations. We contribute our re-implementations of Nearby~\cite{getreuer2018ultrasonic}, PriWhisper~\cite{zhang2014priwhisper}, and Lee et al.'s scheme~\cite{lee2015chirp} in our replication package \cite{replicationpackage}.

\item \textbf{Describe the experimental setup.}  
Clearly specify all system parameters and experimental conditions (e.g., TX/RX devices, volume, gain, payload length, distance, environment, background noise) and include images of the setup. Define the evaluation metrics precisely. Our setup is detailed in \autoref{sec:meta-appendix}.

\item \textbf{Publish data artifacts.}  
Share raw measurement data (such as audio recordings) to enable both replication and reproduction of the evaluation \cite{patil2016statistical}. This transparency allows independent verification and enables other researchers to improve upon the work (e.g., by designing better decoders). We publish a dataset of \num{11900} recorded audio transmissions from eight schemes along with a CSV file containing raw BER results for each recording \cite{replicationpackage}.
\end{itemize}

\subsubsection{Establish standardized testbeds.} For maximum reproducibility, evaluations could be performed in a standardized testbed, such as an anechoic chamber. To simulate realistic conditions, controlled noise sources playing back ambient field recordings or white noise can be introduced.

\subsection{Limitations}\label{sec:limitations}

\subsubsection{Implementation variability}
Due to a lack of research transparency in the field of acoustic data transmission and the absence of publicly available implementations for proposed schemes, we had to re-implement some schemes ourselves.
This process was laborious and prone to errors. Our re-implementations of Nearby \cite{getreuer2018ultrasonic}, PriWhisper \cite{zhang2014priwhisper}, and the scheme by Lee et al.~\cite{lee2015chirp} may differ from the originals because the publications did not provide full implementation details. In part, we had to rely on educated guesses, which we describe in \autoref{sec:reimplementation-challenges}.
Furthermore, for our re-implementations of Nearby and PriWhisper, we used larger message lengths than those reported in the original publications to meet the demands of their intended use cases in a single transmission.
We verified our re-implementations using simulations and practical experiments, which showed that Nearby and PriWhisper can handle these larger message lengths while maintaining high reliability, making them conceptually replicable under these more challenging conditions.
This positive result indicates that our re-implementations did not degrade the performance of the original implementations.
During our experiments, all of our re-implementations were generally robust and reliable for their intended use case.

\subsubsection{Environmental variability}
We conducted our experiments in quiet indoor rooms and a hallway, acknowledging that varying environmental conditions could have influenced our results.
To ensure accuracy, we repeated any experiments that yielded inconsistent results or were affected by unintended noise interference.
However, one measurement stood out: Nearby performed significantly worse in the device models experiment compared to the distance experiment, despite using the same devices and transmission distance.
This was not a measurement error; the only difference was the change in office rooms.
This supports the observation by the Nearby authors, who also observed that the reliability of acoustic data transmissions is sensitive to environmental conditions~\cite[section 8A]{getreuer2018ultrasonic}.
Therefore, given that multipath environments such as office rooms are inherently variable and cannot be precisely replicated by others (see \autoref{sec:severe-multipath}), we also conducted tests in an anechoic chamber, which is a controlled and reproducible testing environment.

\subsubsection{Burst noise variability}\label{sec:limitation-burst-noise}
The burst noise experiment presented several challenges not present in the ambient noise experiment, due to the need for manual handling of the devices. This not only made the process time-consuming---requiring manual effort for 560 transmissions---but also introduced variability in handling, which affected the replicability of the experiment. We attempted to mitigate this by triggering the burst noise at \SI{2}{s} intervals, but some variability remains.
Note that this setup resulted in less burst noise exposure per transmission for the schemes from Gon\c{c}alves et al.~\cite{goncalves2017acoustic} and Lee et al.~\cite{lee2015chirp}, which had shorter transmission durations compared to the other schemes (Gon\c{c}alves et al. at \SI{0.4}{s}, Lee et al. at \SI{1.1}{s}, others approximately \SI{5}{s}; all with a \SI{2}{s} pause between transmissions). Consequently, the results from these two schemes are not directly comparable with those from the others.

\section{Future work}
Based on our evaluation, we have identified a list of design requirements in \autoref{sec:practical-challenges}, which will allow future schemes to be ready for real-world deployment.
In particular, our comparison across multiple smartphones  demonstrated that most schemes struggle to adapt to different device combinations, due to variations in audio hardware such as frequency responses of microphones and speakers, amplifier non-linearities, and potential on-chip post-processing.
Conducting a comprehensive study of acoustic characteristics across a variety of devices, including low-cost \gls{IoT} hardware and smart devices, could delineate the operational limits within which schemes must reliably function. Such knowledge could streamline future testing efforts and assist in selecting relevant test devices.

Additionally, we observed that most prior studies utilized varying experimental setups when testing with real devices, complicating comparisons between results.
Our work highlights the necessity for increased research transparency, specifically in detailing how evaluations are conducted. Future efforts should aim to establish and define a standard for testing acoustic communication schemes.
Our approach, which involved designing a unified experimental setup for multiple schemes, could serve as a foundation (\autoref{sec:meta-appendix}).
However, standardizing testing environments remains a challenge, especially outside of controlled settings like anechoic chambers.

\section{Conclusion}\label{sec:conclusion}
We conducted a systematic literature study on acoustic data transmission schemes and analyzed their generalizability by requesting implementations from authors, re-implementing selected schemes ourselves, and evaluating them on real devices in both realistic and controlled environments.
Our findings reveal that acoustic data transmission schemes often face reliability issues, particularly at higher data rates and across device models.
This problem has not been extensively acknowledged or discussed in previous research, which predominantly relied on theoretical simulations or conducted only limited real-device testing.
Moving forward, we recommend stronger emphasis on evaluating real-world performance, and we offer our evaluation methodology as a foundation for future work.

\textbf{RQ1.} \textit{``How challenging is it to obtain or re-implement systems proposed in this field?''}
The state of replicability in this field is bleak, as not a single publication in our systematic literature study provided accessible implementations. Contacting authors for their implementations proved largely unsuccessful, and re-implementating these systems is both time-consuming and challenging due to the inconsistent information provided in the publications.

\textbf{RQ2.} \textit{``How well do these schemes generalize to ad-hoc smartphone communication use cases?''}
Most of the schemes we tested are not yet reliable for practical ad-hoc data transmission between nearby smartphones. Although many claim high throughput, they fail to sustain it in real conditions.
However, the scheme by Lee et al.~\cite{lee2015chirp} proved reliable even at longer distances, and the non-academic ggwave~\cite{gerganov2024ggwave} is effective for close to medium distances.

\textbf{RQ3.} \textit{``Which practical challenges affect nearby acoustic data transmission between smart devices?''}
Acoustic data transmission on smart devices faces additional, often overlooked limitations compared to RF communication, including severe inter-symbol interference from long delay spreads and the added complexity of non-Gaussian noise and device diversity.
Because simulations cannot accurately capture these real-world complexities, we systematically tested acoustic schemes across multiple device models under varying environmental and noise conditions to quantify their impact on reliability and generalizability. 
Our findings show that thorough practical testing  clarifies the trade-off between throughput and reliability for specific use cases, guiding the selection of an optimal system for practical applications.
To achieve practical applicability, future schemes should be resistant to delay spreads of tens of milliseconds, withstand large volume fluctuations due to device heterogeneity, and handle both continuous and impulsive noise.

\section*{Availability}\label{sec:availability}
To provide transparency about our research results, we provide a replication package \cite{replicationpackage} containing:

\begin{itemize}
    \item A Methodological Transparency \& Reproducibility Appendix  (META) containing detailed descriptions of our experimental setup, including the exact settings for transmission and reception (see \autoref{sec:meta-appendix}).
    \item Our three re-implementations of acoustic communication schemes \cite{getreuer2018ultrasonic,zhang2014priwhisper,lee2015chirp} in MATLAB (see \autoref{sec:obtaining-implementations}).
    \item A collection of \num{11900} WAV files with recorded transmissions from eight acoustic communication schemes, captured both in an office environment and an anechoic chamber (see \autoref{sec:evaluation}).
    \item A CSV file with our raw data, including the BER for each scheme's recordings.
    \item The full MATLAB analysis code that decodes recordings from different schemes in parallel, along with R scripts that generate all result figures.
\end{itemize}

\begin{acks}
We thank all authors who replied to us when we requested their acoustic data transmission implementations following our literature study (\autoref{sec:obtaining-implementations}).
Furthermore, we thank the anonymous reviewers for their helpful suggestions.
This work has been funded by the LOEWE initiative (Hesse, Germany) within the emergenCITY center [LOEWE/1/12/519/03/05.001(0016)/72].
\end{acks}

\bibliographystyle{ACM-Reference-Format}
\typeout{}
\bibliography{bibliography}

\appendix

\section{Methodological transparency \& reproducibility appendix}\label{sec:meta-appendix}

This section offers detailed information about our experimental setup with the goal of improving the transparency and replicability of our work. We document the exact settings we used for transmission and reception. \autoref{tab:hardware-software} lists all the hardware and software we used, including version numbers.

\subsection{General}
For every experiment, we repeated the measurements in case of external effects, such as vocal or noise interference, or when there were problems with playback and recording of the transmission.

\subsection{Transmission}
Following the setup shown in \autoref{fig:experimental_setup}, we used the smartphones listed in \autoref{tab:hardware-software}. First, we generated transmission WAV files for each scheme, which are part of our dataset.
Message lengths were set to the publication's default or if the implementation supported custom message sizes we adjusted it for an approximate five-second transmission, which we consider a practical upper limit for user wait times.
All audio signals were then normalized by dividing the entire signal by the highest absolute sample value and scaling it to \SI{-3}{dBFS}.

We transmitted these files on each smartphone using the open-source Android app \textit{Vinyl}, which allows automating the playback of multiple WAV files using playlists. We compiled a playlist containing the transmission files from all schemes, separated by two-second-long silent WAV files to prevent reverberation from affecting subsequent transmissions.
For convenience, we also provide a single WAV file containing repeated transmissions from all schemes and corresponding silent gaps for easy playback.
We controlled the output volume using the open-source Android app \textit{Volume Control}, which provides precise control of the Android volume index, as shown in \autoref{tab:device-models-smartphones}.

\subsection{Reception}
On the receiver smartphone, we recorded each transmission using the open-source Android app \textit{Audio Recorder}, which allows detailed control over the microphone source. We selected the \texttt{MIC} source for all smartphones, except for the Google Pixel 4a, which also supported the \texttt{UNPROCESSED} source to bypass Android's post-processing. 
Recordings were saved as \SI{24}{bit} PCM float mono WAV files at a minimum of \SI{44.1}{kHz}, with software gain at 100\% and all extra filtering disabled.

We recorded multiple transmissions consecutively and later segmented them manually in the \textit{REAPER} digital audio workstation.
To streamline this process, we used a script\footnote{REAPER script for splitting: \url{https://stash.reaper.fm/v/25677/SplitX.lua}} for batch cutting and a custom iterator for unique file naming. The segments were rendered as \SI{32}{bit} PCM mono WAV files at a minimum of \SI{44.1}{kHz}.
Finally, we decoded these files with each scheme's receiver implementation to calculate the TER. If a decoded message was longer than expected, we truncated it; if it was shorter, we counted the missing bits as errors.

\subsection{Experimental setup}
We conducted tests in several indoor locations and an anechoic chamber. For the experiments in the indoor environment, we mounted all smartphones on small tripods with rubber feet, which isolated them from vibrations caused by the their speakers.
We positioned the smartphones so their speakers and microphones faced directly toward each other (on-axis), measuring the distance between.
The dimensions (length x width x height) of the indoor spaces were: \textit{small office} (\SI{5.2}{m} x \SI{4}{m} x \SI{3}{m}), \textit{large office} (\SI{13.5}{m} x \SI{5.3}{m} x \SI{2.8}{m}), \textit{meeting room} (\SI{7.9}{m} x \SI{5.3}{m} x \SI{3}{m}), \textit{lecture room} (\SI{13.5}{m} x \SI{6.6}{m} x \SI{3.6}{m}), and the \textit{hallway} for long-distance tests (\SI{40}{m} x \SI{2}{m} x \SI{3}{m}).

The \textit{anechoic chamber} (\SI{5.2}{m} x \SI{7.7}{m} x \SI{5.8}{m}) provided a reflection-free environment, 
with walls lined with mineral wool cones coated in synthetic resin (cone length \SI{100}{cm}, cone base area \SI{24}{cm} x \SI{24}{cm}, reflection coefficient below 0.01 for frequencies above \SI{100}{Hz}).

\subsubsection{Preliminary best-case setup}\label{sec:experiment-basic-success-setup}

For the preliminary tests (\autoref{sec:experiment-basic-success}), we established a best-case baseline using studio equipment instead of smartphones. For transmission, a Neumann KH 80 DSP studio speaker played the signals (local control off, acoustic control ``free-standing'', output level \SI{94}{dBSPL}, input gain at \SI{0}{dB}). We connected the speaker to an RME ADI-2 Pro FS R BE audio interface (\SI{-40}{dB} gain at a reference level of \SI{+24}{dBu}).

For recording, a calibrated Earthworks M23R reference microphone captured the audio. The signal was fed through a Camden EC1 preamp (\SI{+35.5}{dB} gain, no additional filtering) into the audio interface (without additional gain). 
The preamp's output was split to both interface inputs and summed to mono using mid-side processing, increasing the SNR by \SI{3}{dB}.
We used REAPER for playback and recording.
The speaker and microphone were aligned on-axis at a distance of \SI{50}{cm}.

\subsubsection{Ambient noise setup}\label{sec:experiment-noise-setup}

For the ambient noise experiment (\autoref{sec:experiment-noise}), we used the standard smartphone and introduced noise using the Neumann KH 80 DSP studio speaker. The speaker was placed \SI{1}{\meter} from the receiver, orthogonal to the transmission axis. 
The speaker configuration was the same as in the preliminary tests (\autoref{sec:experiment-basic-success-setup}), with the exception of the output gain of our interface, which was set to \SI{-35}{dB} to subjectively match the real-life noise levels of the scenarios at the receiver's location.
During each iteration of the experiment, one field recording was played repeatedly while the transmission sounds were played sequentially. This process was repeated for all three recordings (café, train station, marketplace), which are freely available online\footnote{Café field recording (CC BY 3.0): \url{https://freesound.org/people/AshFox/sounds/172968/}, train station field recording (CC BY 4.0): \url{https://freesound.org/people/Kyster/sounds/121576/}, marketplace field recording (CC BY 3.0): \url{https://freesound.org/people/le_abbaye_Noirlac/sounds/129677/}} under a CC-BY license.

\subsubsection{Dynamic environment setup}\label{sec:experiment-environment-setup}

To evaluate performance in a time-varying channel (\autoref{sec:experiment-environment}), we conducted tests in a lecture room with four people walking slowly and quietly (without talking) around the devices. The smartphones were mounted on small tripods placed on tables in the center of the room, 50 cm apart on-axis. For comparison, a static control test was performed with an identical furniture layout but without people.

\subsection{Replicating our experiments}
The procedures described above should allow for the replication of our experiments. However, the reliability of acoustic transmissions is sensitive to factors like specific device models, microphone/speaker quality, and environmental acoustics.
While we expect the overall trends to be similar, absolute TER values will likely differ.

\begin{table}[bt]
\centering
\caption{\captionheadline{Hardware and software used during our experiments.}}
\label{tab:hardware-software}
\small
\begin{threeparttable}
\begin{tabular}{l lll}
\toprule
\textbf{Type} & \textbf{Manufacturer} & \textbf{Product} & \textbf{Version} \\ 
\toprule
Microphone & Earthworks & M23R& \\
Preamp & Camden & EC1 & \\
Audio Interface & RME & ADI-2 Pro FS R BE\\
Calibration Device & Galaxy Audio & CM-C200 &\\
Loudspeaker & Neumann & KH80 DSP &\\
\midrule
Smartphone & Google & Pixel 4a & 128GB\\
Smartphone & Google & Pixel 6 Pro & 5G Sub-6 128GB\\
Smartphone & Huawei & Nexus 6P & 32GB\\
Smartphone & Oppo & Reno 6 & Dual-SIM 5G 128GB\\
Smartphone & Samsung & Galaxy S20 Ultra & 5G 128GB\\
\midrule
Desk Microphone Stand & K\&M & 23150-3 & \\
Ball Joint & Roadworx & Universal Ball Joint & \\
Smartphone Clamp & Roadworx & Smartphone Clamp & \\
\midrule
Smartphone App & Adrien Poupa & Vinyl Music Player\tnote{1} & v1.5.0\\
Smartphone App & Stas Shakirov & Volume Control\tnote{2} & v2.6.0\\
Smartphone App & axet & Audio Recorder\tnote{3} & v3.5.15\\
\midrule
DAW & Cockos & REAPER\tnote{4} & v6.82\\
\bottomrule
\end{tabular}
  \scriptsize
  \begin{tablenotes}
    \item[1] Vinyl: \url{https://f-droid.org/en/packages/com.poupa.vinylmusicplayer/}
    \item[2] Volume Control: \url{https://f-droid.org/packages/com.punksta.apps.volumecontrol/}
    \item[3] Audio Recorder: \url{https://f-droid.org/en/packages/com.github.axet.audiorecorder/}
    \item[4] REAPER: \url{https://www.reaper.fm/}
  \end{tablenotes}
\end{threeparttable}
\end{table}

\section{Re-implementations}\label{sec:reimplementation-challenges}

In this section, we detail the challenges we encountered during the re-implementation of Lee et al.'s scheme, Nearby, and PriWhisper.

\subsection{Lee et al.}

We were able to implement all aspects of this scheme es described in their paper~\cite{lee2015chirp}.
Their practice of listing most system parameters clearly in a table was particularly helpful. 
We interpret the \SI{16}{bps} data rate mentioned by Lee et al.~as the gross data rate, as it appears they did not account for the preamble in their calculations. Based on the default system parameters provided, we calculate a net data rate of \SI{14.55}{bps}.

\subsection{Nearby}
Implementing the transmitter for the Nearby system~\cite{getreuer2018ultrasonic} was straightforward, following the information provided in the paper. However, re-implementing the receiver proved challenging, as some aspects of the paper were unclear to us.

\subsubsection{Synchronization}
To locate the transmission's start within the recording, the system processes the signal by iterating over samples, segmenting the signal from each sample offset, and correlating these segments with a period of the code signal.
This correlation indicates the likelihood that a symbol begins at each sample offset, as each MFSK symbol was spread by the code signal. After correlating across all time offsets, the results are normalized to compensate for any potential Doppler shifts. These results are termed ``normalized acquisition scores'' in their paper (Section VI.C in \cite{getreuer2018ultrasonic}).
Next, we identify the spacer symbol's position using the normalized acquisition scores.
Ideally, a distinct peak at the spacer symbol’s position would appear when plotting these scores. However, since the same code signal spreads all token symbols, peaks appear for each symbol, not just the spacer. We implemented a method to identify the positions of the $n$ highest peaks---where $n$ represents the number of token symbols---and select the earliest peak as the start. The paper does not discuss this issue or how to resolve it.

\subsubsection{Block processing}
The authors mention that their implementation processes signals in \SI{100}{ms} blocks to likely improve time efficiency, yet the paper does not detail how this feature is implemented.
We chose not to implement this aspect, as it is unlikely to impact our evaluation results.

\subsubsection{Ambiguity in Algorithm 2}
Algorithm 2 in the paper \cite{getreuer2018ultrasonic} employs a counter variable $n$ in line 2 that initializes at zero and increments by one in each iteration to compute a ``raw acquisition score'' for each sample offset $n$, representing the likelihood of a data symbol starting at that offset. However, the authors do not specify an upper bound for $n$. We have documented our method for determining this upper bound in the comments of our receiver function, although it remains unclear  whether this method aligns with the authors' intended approach.

\subsubsection{Ambiguity in Equation 47}
In Equation 47 in the paper \cite{getreuer2018ultrasonic}, the variable $c$, representing the sinc-interpolated code sequence, is indexed at position $(kM_p+m)/F_b$. Here, $F_b$ denotes a sample rate of \SI{12}{kHz}, which means the division results in a time in seconds at which the code signal should be indexed.
Since $c$ is a discrete signal sampled at \SI{48}{kHz}, it cannot be indexed at arbitrary times.
To convert the time $(kM_p+m)/F_b$ to a corresponding sample offset in $c$, we multiply by \num{48000}, the sample rate. 
Again, it is not clear if the authors intended this.

\subsubsection{Ambiguity in Algorithm 3}
Algorithm 3 in the paper \cite{getreuer2018ultrasonic} uses the matrix of raw acquisition scores $a^{raw}$, generated by Algorithm 2. In line 8 of Algorithm 3, it accesses row $n-d$ of $a^{raw}$, where $n$ is a counter variable starting at zero and increasing by one each iteration. The paper does not specify an upper bound for $n$, but we assume that $n$ runs up to the number of rows in $a^{raw}$ reduced by one.
The variable $d$ is set to 215, leading to the algorithm attempting to access negative row indices during the first 215 iterations, which is invalid.
We resolved this issue by omitting line 8 in iterations where $n-d$ results in a negative index. However, as a consequence, the normalized acquisition score matrix ends up with 215 fewer rows than $a^{raw}$. It is unclear whether this reduction is in line with the authors' intentions.

\subsubsection{Ambiguity during noise suppression}
The bandpass filter for noise suppression uses a variable $\eta$, which is supposed to be ``a small positive parameter'' (Section VI.A in \cite{getreuer2018ultrasonic}). Since the paper does not specify an exact value for $\eta$, we set it to 1.

\subsubsection{Data rate}
We interpret the stated data rate of \SI{94.5}{bps} as the gross data rate. The system can operate without the token structure that includes additional spacers and a parity symbol, in which case \SI{94.5}{bps} is also the net data rate. However, when using the full token structure as described in the paper and used in our evaluation, the net data rate decreases to \SI{84}{bps}.

\subsection{PriWhisper}

When re-implementing PriWhisper~\cite{zhang2014priwhisper}, the major difficulty was designing a mechanism for synchronization and calibration, as the paper does not clearly explain these aspects. Additionally, the descriptions of the interleaving and checksum processes in the paper were ambiguous.
The specified data rate was also unclear, further complicating our implementation efforts.

\subsubsection{Synchronization}
In the original implementation, a sinusoid preamble is used within the jamming signal to aid synchronization. However, for our study, we did not implement the security features of the system, focusing solely on data transmission. 
To minimize the impact on the data rate, we combine the synchronization and calibration sequence.
Accordingly, we introduced a special symbol at the beginning of each transmission, which contains the sum of all available data waveforms $s(t)$ (Section III.A in \cite{zhang2014priwhisper}).
For synchronization, we use this known synchronization symbol and cross-correlate it with the recording. The position of the highest peak in the output is considered the start of the transmission.

\subsubsection{Calibration}
To determine the frequency of an MFSK symbol, we correlate each symbol duration in the recording with all available frequencies using the following equation from Section III.A in \cite{zhang2014priwhisper}:
\begin{equation}\label{eq:priwhisper:demodulation}
	R_m = \left|\int_{0}^{T}r(t)e^{i2\pi(f_c + m\Delta f)t}dt\right|, m\in[0, M-1]
\end{equation}
After iterating over all frequencies, we select the frequency with the highest correlation value and map it to the binary number it represents. We repeat these steps for all symbols to decode the entire data transmission in the recording.
Before demodulation, calibration using the synchronization symbol is crucial due to potential variations in microphone sensitivity across frequencies. Without calibration, \autoref{eq:priwhisper:demodulation} might return a lower correlation value for the actual frequency present in a symbol due to the frequency-selectivity of the microphone.

To address this, we apply \autoref{eq:priwhisper:demodulation} to the synchronization symbol, containing all frequencies. This establishes a baseline correlation for each frequency. We then calculate a calibration factor for each frequency by dividing the highest of these baseline values by the others. This factor adjusts the correlation values during demodulation, ensuring that each frequency is appropriately represented despite the microphone's bias.
However, this calibration step assumes that the channel characteristics remain constant throughout the transmission duration.

\subsubsection{Interleaving and checksum}
The paper mentions interleaving the data and adding a checksum, but it lacks clear guidance on how to implement the interleaving process and the sequence of these steps. In our implementation, we first perform regular random interleaving and then add the checksum.

\subsubsection{Data rate}\label{sec:priwhisper-data-rate}
We calculate a gross data rate of \SI{771}{bps} for $M=8$, which differs from the \SI{1027}{bps} reported by the authors.
Our data rate calculation is as follows: Using $M$ different frequencies allows encoding $log_2(M)$ bits per symbol. For $M=8$, this results in 3 bits per symbol. With a symbol duration of \SI{2}{ms}, we transmit \num{500} symbols pers second, equivalent to \num{1500} bits per second. As one coded data block contains \num{255} BCH encoded bits, we can transmit $\frac{1500}{255} \approx 5.88$ blocks per second. One block contains \num{131} bits of pure data, resulting in a gross data rate of $5.88 \times 131 \approx \SI{771}{bps}$, excluding signal components for synchronization or calibration. The resulting net data rate is \SI{729}{bps}.

\end{document}